\documentclass[12pt,a4paper]{article}
\usepackage{color,graphics,amsmath,epsfig,rotating}
\usepackage{axodraw}
\usepackage{pstricks}

\begin{document}

\pagestyle{plain}
\bibliographystyle{plain}

\setlength{\unitlength}{1mm}
\renewcommand{\arraystretch}{1.4}


\def\mxs{M_{X_S}}
\def\cnur{c_{\nu_R}}
\def\ctau{c_{\tau_R}}
\def\cnul{c_{\tilde\nu'_L}}
\def\nul{\tilde\nu'_L}
\def\ctl{c_{t_L}}
\def\ctr{c_{t_R}}
\def\cbr{c_{b_R}}
\def\ccl{c_L}
\def\ccr{c_R}
\def\ctaul{c_{\tau_L}}
\def\mzp{M_{Z'}}
\def\mlzp{M_{LZP}}
\def\sw{s_W}
\def\cw{c_W}
\def\gten{g_{10}}
\def\mnur{M_{\nu_R}}
\def\mnup{M_{\nu'}}
\def\nur{\nu_R}
\def\nup{\nu^{\prime}}
\def\gz{g_Z}
\def\gzp{g_{Z'}}
\def\mnul{M_\nul}
\def\gzzp{g_{ZZ'}^{\nur}}
\def\gzlr{g_Z^{\nur\nul}}
\def\mixzzp{\theta_{ZZ'}}
\def\tb{\tan\beta}
\def\x{\chi}
\def\ti{\tilde}
\def\noi{\noindent}
\def\mnul{M_{\nu'_L}}
\def\gz{g^{\nu_R}_Z}

\def\sigv{\langle\sigma v\rangle}

\def\micromegas{{\tt micrOMEGAs\,2.0}}
\def\micro{{\tt micrOMEGAs}}
\def\calchep{{\tt CalcHEP}}

\newcommand{\beqn}{\begin{eqnarray}}
\newcommand{\eeqn}{\end{eqnarray}}

\newcommand{\eq}[1]  {\mbox{(\ref{eq:#1})}}
\newcommand{\fig}[1] {Fig.~\ref{fig:#1}}
\newcommand{\Fig}[1] {Figure~\ref{fig:#1}}
\newcommand{\tab}[1] {Table~\ref{tab:#1}}
\newcommand{\Tab}[1] {Table~\ref{tab:#1}}

\newcommand{\gsim}{\;\raisebox{-0.9ex}
           {$\textstyle\stackrel{\textstyle >}{\sim}$}\;}

\newcommand{\lsim}{\;\raisebox{-0.9ex}{$\textstyle\stackrel{\textstyle<}
           {\sim}$}\;}

\newcommand{\smaf}[2] {{\textstyle \frac{#1}{#2} }}
\newcommand{\sfrac}[2] {{\textstyle \frac{#1}{#2}}}
\newcommand{\pif}      {\smaf{\pi}{2}}

\def\delr            {\!\stackrel{\leftrightarrow}{\partial^\mu}\!}

\newcommand{\bb} {\color{blue}}
\newcommand{\bbf} {\color{blue}\bf}
\newcommand{\change} {\bf\mbf}


\begin{flushright}
   \vspace*{-18mm}
   LAPTH-1184-07\\
   CERN-PH-TH/2007-083
\end{flushright}
\vspace*{2mm}

\begin{center}

{\Large\bf Dirac  Neutrino  Dark Matter} 

\vskip 10pt

{\large   Genevi\`eve B\'elanger$^a$, Alexander Pukhov$^b$ and G\'eraldine Servant$^{c,d}$}

\vskip 20pt

\centerline{$^{a}$ {\it Laboratoire de Physique Th\'eorique LAPTH, F-74941 Annecy-le-Vieux, France}}
\vskip 2pt
\centerline{$^{b}$ {\it Skobeltsyn Inst. of Nuclear Physics, Moscow State Univ., Moscow
119992, Russia}}
\vskip 2pt
\centerline{$^{c}$ {\it CERN Physics Department, Theory Division, CH-1211 Geneva 23, Switzerland}}
\vskip 2pt
\centerline{$^{d}$ {\it Service de Physique Th\'eorique, CEA Saclay, F91191 Gif--sur--Yvette,
France}}
\vskip 3pt
\centerline{\tt  belanger@lapp.in2p3.fr, pukhov@lapp.in2p3.fr, geraldine.servant@cern.ch}


\end{center}

\vskip 13pt

\begin{abstract}
\vskip 3pt
\noindent

We investigate the possibility that dark matter is made of heavy Dirac neutrinos with mass $m \in [{\cal O}(1)$ GeV-- a few TeV] and with suppressed but non-zero coupling to the Standard Model $Z$ as well as a coupling to an additional $Z'$ gauge boson.  The first part of this paper provides a model-independent analysis for the relic density and direct detection in terms of  four main parameters: the mass, the couplings to the $Z$, to the $Z'$ and to the Higgs.
These WIMP candidates arise naturally as Kaluza-Klein states in extra-dimensional models with extended electroweak gauge group $SU(2)_L\times SU(2)_R \times U(1)$.
They can be stable because of Kaluza-Klein parity 
or of other discrete symmetries related to baryon number for instance, or  even, in the low mass and low coupling limits, just because of a phase-space-suppressed decay width.
 An interesting aspect of warped models is that the extra $Z'$ typically couples only to the third generation, thus avoiding the usual experimental constraints. 
 In the second part of the paper, we illustrate the situation in details in a warped GUT model.

\end{abstract}
\vskip 13pt
\newpage

\section*{Introduction}

It is well-known that a heavy ($m\gsim 1$ GeV) Dirac Neutrino with  Standard Model interactions  is ruled out as dark matter 
because of its large coupling to the $Z$. On the one hand, it annihilates too strongly into $Z$ to have the right thermal
 abundance. On the other hand, even if it had the right relic density from a non-standard production mechanism, 
it would scatter elastically off nuclei with a large cross section induced by the $Z$ exchange and should have 
been seen in direct detection experiments, unless its mass is larger than several tens of 
TeV \cite{Servant:2002hb,Goodman:1984dc}. Moreover, this type of neutrino is excluded by electroweak precision tests \cite{Yao:2006px}.
In contrast, a sterile Majorana neutrino as dark matter is a possibility that has raised interest lately
\cite{Dodelson:1993je,Asaka:2005an}. In this case, the neutrino mass $m_s$ that has been considered is rather in the keV- MeV range and behaves as warm dark matter if $m_s\lsim 20$ keV. In addition, the Majorana mass scale  is determined by the see-saw formula for neutrino masses.

In the present work, we are considering a different type of Dirac neutrino, denoted $\nup$, corresponding to a typical cold dark matter WIMP candidate with a mass at the electroweak scale, suggesting that dark matter and the electroweak scale are somehow related (even if the mass of the dark matter particle does not come from electroweak symmetry breaking). We do not assume any particular relation  between the mass scale of $\nup$ and that of the light standard model neutrinos.
The relic density of $\nup$ is entirely determined by the standard thermal mechanism and therefore by its annihilation cross section.
We are assuming that the coupling to the Standard Model $Z$ is suppressed. There are various reasons why this can happen. A typical framework is to start  with an $SU(2)_L$ singlet neutrino but charged under $SU(2)_R$. 
 Because the gauge bosons of $SU(2)_R$ are heavy, their interactions with $\nu_R$ are quite feeble and this makes $\nur$ behave as a WIMP.
In addition,  electroweak (EW) symmetry breaking typically induces a mixing between $Z$ and $Z'$, leading to an effective small coupling of $\nu_R$ to the $Z$.
Examples of this type were studied in warped extra dimensions \cite{Agashe:2004ci,Agashe:2004bm,Hooper:2005fj}, and in universal extra dimensions \cite{Hsieh:2006qe}.
Note that the most promising realistic warped extra-dimensional scenarios need the EW gauge group to be extended to $SU(2)_L\times SU(2)_R \times U(1)$. In this context, Kaluza-Klein Dirac neutrinos charged under $SU(2)_R$ are necessarily part of these constructions, even though their stability typically requires an additional ingredient. For instance, it was shown in \cite{Agashe:2004ci,Agashe:2004bm} that
implementing baryon number conservation in warped GUTs leads to the stability of a KK RH neutrino.

Finally, even in the absence of an additional symmetry, $\nup$ can be cosmologically stable if the couplings involved in its decay are very suppressed. This can happen even if the neutrino has a large annihilation cross section providing the correct relic density.

Dirac neutrino dark matter was also studied in the 4D models of Ref.~\cite{Schuster:2005ck} where the $Z$-coupling suppression has a different origin and results from mass mixing between gauge eigen states with opposite isospin. Ref.~\cite{Schuster:2005ck} did not consider the effect of a $Z'$ that we study in details here.  
In the first part of this paper, we present a model-independent analysis for the viability of Dirac Neutrino dark matter in 
terms of three main parameters, the mass, the $Z$ coupling and the $Z'$ coupling of $\nup$. We also discuss the effect of a 
coupling with the Higgs. The remaining part of the paper is a refined analysis of the dark matter 
candidate which arises in the warped GUT models of Ref.~\cite{Agashe:2004ci,Agashe:2004bm}.
In both studies, the computation of the relic density is performed with \micro 2.0~\cite{Belanger:2006is}
after implementing  new model files into \calchep~\cite{Pukhov:2004ca}.

\vskip 3cm
{\Huge{\bf Part I -  Model-independent analysis}}\\

In this part, we  consider a generic extension of the Standard Model (SM)
containing a stable heavy Dirac  neutrino denoted $\nup$, with mass $\mnup$.   
We assume a parity symmetry (except in Section \ref{subsec:lightneutrino}) that does not act on SM fields and $\nup$ is the lightest new particle charged under it.
The model also contains an additional $Z'$ gauge boson and potentially a charged gauge boson $W'$, with masses $M_{Z'},\ M_{W'}\gsim 500$ GeV.
To avoid most low
energy constraints we will assume that these new gauge bosons couple
only to the fermions of the third generation. 
 $\theta_{ZZ'}$ is the mixing  which induces
the  $Z'WW$ and ${Z'ZH}$  couplings.
We also introduce $g_H$ as the Higgs
coupling to the $\nup$ (for instance induced via mixing with a heavy $\nu_L$ as illustrated in Part II).
We assume that only one chirality of $\nup$ couples to the gauge bosons (in our numerical examples, we chose the right-handed chirality).
The effective couplings of
$\nup$ to $Z$,  $Z'$ and $H$ are denoted $g_{Z}$, $g_{Z'}$ and $g_H$ respectively:
\begin{equation}
g_Z\overline{\nup}\gamma^{\mu}\frac{1+ \gamma_5}{2} \nup Z_{\mu} \ ,  \ \ \ \ \  g_{Z'}\overline{\nup}\gamma^{\mu}\frac{1+ \gamma_5}{2} \nup Z'_{\mu} \  , \ \ \ \ \ g_H\overline{\nup} \nup H
\end{equation}
We work at the level of a low energy effective theory. We assume that the remaining new physics which makes the model more complete does not interfere much with our dark matter analysis.

\section{Direct detection constraints}

\begin{figure}[!htb]
  \centerline{\epsfig{file=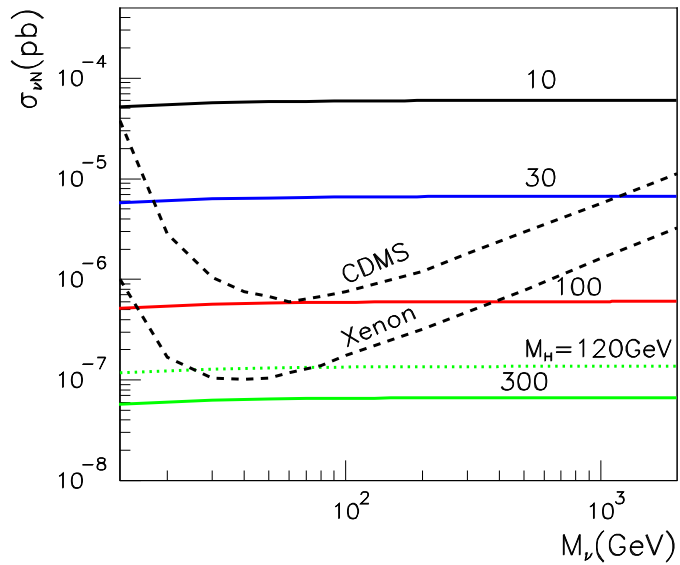, width=11cm,height=10.75cm}}
   \caption{Neutrino-neutron scattering cross section due to $Z$-exchange  
   for  $g/g_z=10,30,100,300$ where  $g=e/(\sin \theta_W)$ is the SM coupling. 
The dotted line shows the effect of adding the Higgs exchange for $g_H$=0.25, $m_H=120$ GeV, in the case where $g/g_z=300$. Also represented is the CDMS limit as well as the recent XENON limit \cite{latest_xenon}. 
 } \label{fig:directdetection}
\end{figure}

Direct detection constraints are very simple for a Dirac neutrino (see for instance section 3 of Ref.~\cite{Servant:2002hb}). The cross-section $\sigma_N$ for $\nup$ scattering on nucleons is governed by the $t$-channel exchange of the $Z$,  the $Z'$ exchange is comparatively negligible. 
In contrast with Majorana dark matter, the $Z$ exchange contributes to the spin-independent scattering cross section.
Therefore, strong constraints on $g_Z$, the $\nup$ coupling to the $Z$, are derived from direct detection experiments, in particular CDMS \cite{Akerib:2004fq}  and  recently  XENON which has now the most stringent limit \cite{latest_xenon}.
The dependence on $\mnup$ that we observe for this
constraint is related to the experimental sensitivity,
which is optimal around 50--100 GeV. The theoretical prediction does
not depend on $\mnup$, as illustrated on Fig.~\ref{fig:directdetection}.
In our plots, we use the parameter $g/g_Z$ where $g$ is the SM electroweak coupling $g=e/(\sin \theta_W)$. 
The direct detection limit  has been rescaled to take into
account the fact that, for a
Dirac fermion, the interaction of the $Z$ with protons is suppressed by a factor $(1-4 \sin^2 \theta_W)^2$ (see Eq.~\ref{eq:spin_dependent}) so that  the scattering with nucleons is completely dominated by neutrons. When CDMS and XENON
quote their limit they rather assume that protons and neutrons
contribute equally. This means that the CDMS and XENON exclusion curves go
up by a factor $A^2/((1-4\sin^2\theta_W)Z-(A-Z))^2$ i.e 3.7 and 3.4 respectively. 
We should keep in mind  that the bound from direct detection experiments is subject to some
astrophysical uncertainties such as the velocity
distribution of the WIMP. Based on Ref.\cite{Bottino:2005qj} we could estimate these
uncertainties and allow for a factor $\sim 3$ in the interpretation of the CDMS and XENON limits, over the full 10 GeV - 1 TeV WIMP mass range even though there are actually much larger uncertainties for masses below $\sim$ 40 GeV. For clarity, we have not displayed this uncertainty in our plots. 

It is clear from Fig.~\ref{fig:directdetection} that we have to impose  $g_Z \lsim  g/100$  for $\mnup \lsim $ 400 GeV to satisfy the XENON constraint. 
If $\nup$ has a sizable coupling to the Higgs, the elastic scattering via Higgs exchange is not always negligible compared to the 
$Z$-exchange especially when $\nup$ couples weakly to the $Z$. The spin-independent elastic scattering cross section on nucleons is the sum of two contributions (when averaging over $\nup$ and $\overline{\nup}$, the negative interference term cancels):
\begin{equation}
\sigma_{\stackrel{neutron}{proton}}=\frac{g^2 m_{n/p}^2 }{64\pi M_Z^4 \cos^2 \theta_W} \left[ g_Z^2  \left(\stackrel{1}{(1-4\sin^2\theta_W)^2}\right)+\left(\sum_q f_{T,q}^{n/p}  4 \cos \theta_W
g_H \frac{m_{n/p}}{M_W}\frac{M_Z^2}{M_H^2}\right)^2\right]
\label{eq:spin_dependent}
\end{equation}
where $f_{T,q}^{n,p}$ are taken from Ref.~\cite{Jungman:1995df}.
For example, when $m_H=120$ GeV, $g_H=0.25$ and $g/g_Z=300$,  the Higgs and $Z$ 
contributions become comparable for elastic scattering on neutrons, as
illustrated in Fig.~\ref{fig:directdetection}, while  for elastic scattering on protons, 
the Higgs contribution dominates.

In contrast with Majorana fermions like in the MSSM, the parameter that determines the elastic scattering of Dirac neutrino dark matter on nucleus is the same parameter that drives annihilation and determines the relic density. We now look at when Dirac neutrinos can  inherit the  correct thermal abundance.
\section{Annihilation}
\begin{figure}[!htb]
 \centerline{\epsfig{file=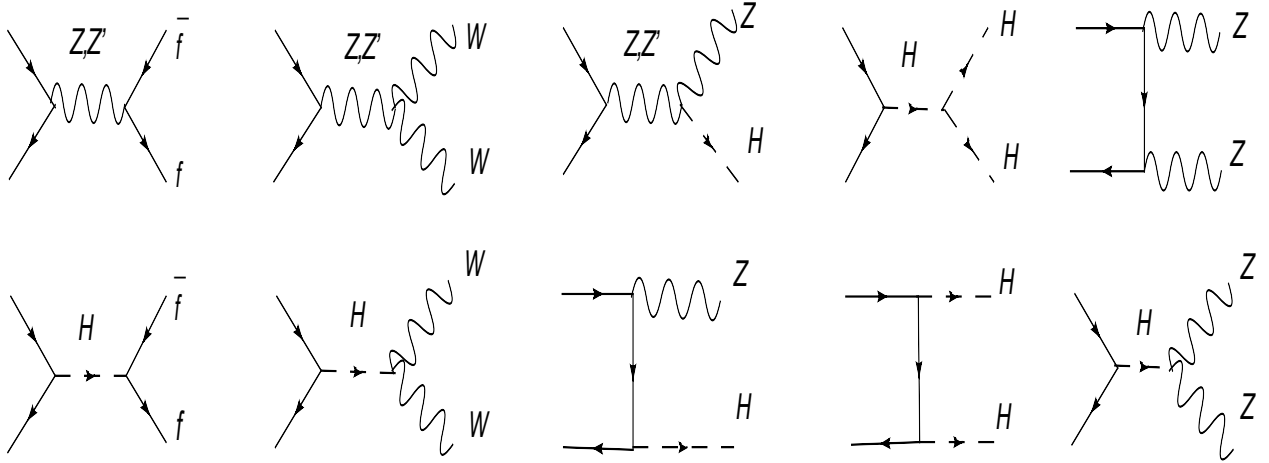, width=16.5cm,height=6.5cm}}
  \caption{ $\nup$ annihilation diagrams into $f\overline{f}$, $WW$, $Zh$, $hh$ and $ZZ$.} \label{fig:Feyn_Diag}
\end{figure}
\subsection{Annihilation via s-channel $Z$ exchange}
\begin{figure}[!htb]
  \centerline{\epsfig{file=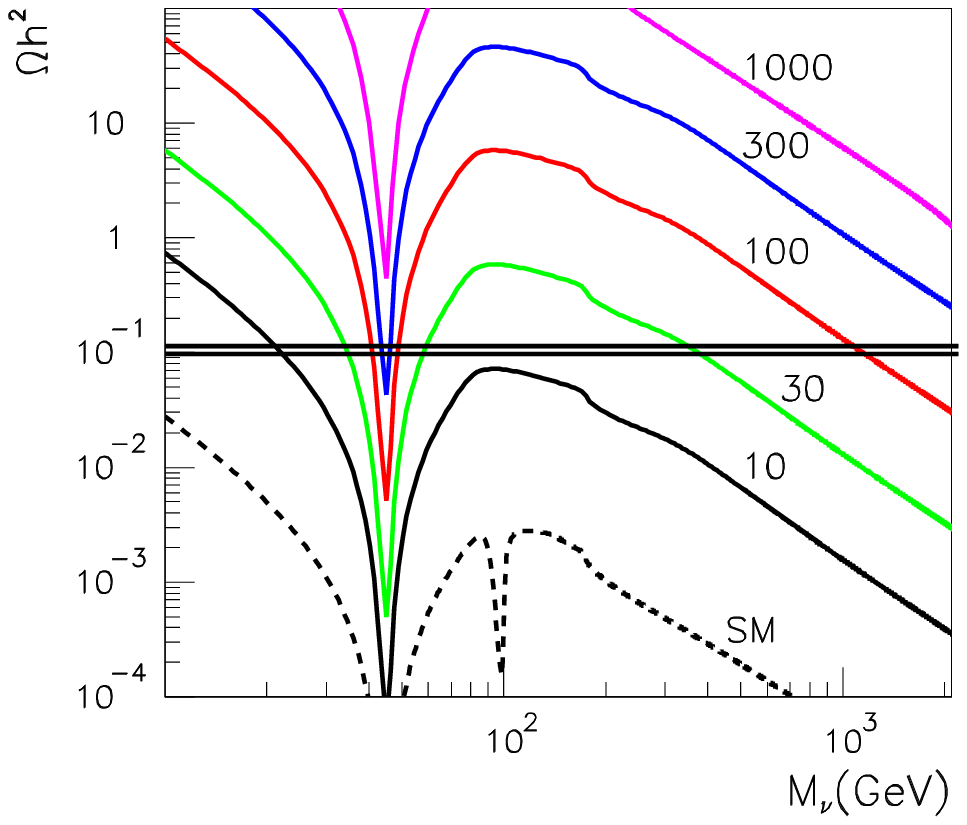, width=9.5cm,height=9.5cm}
  \epsfig{file=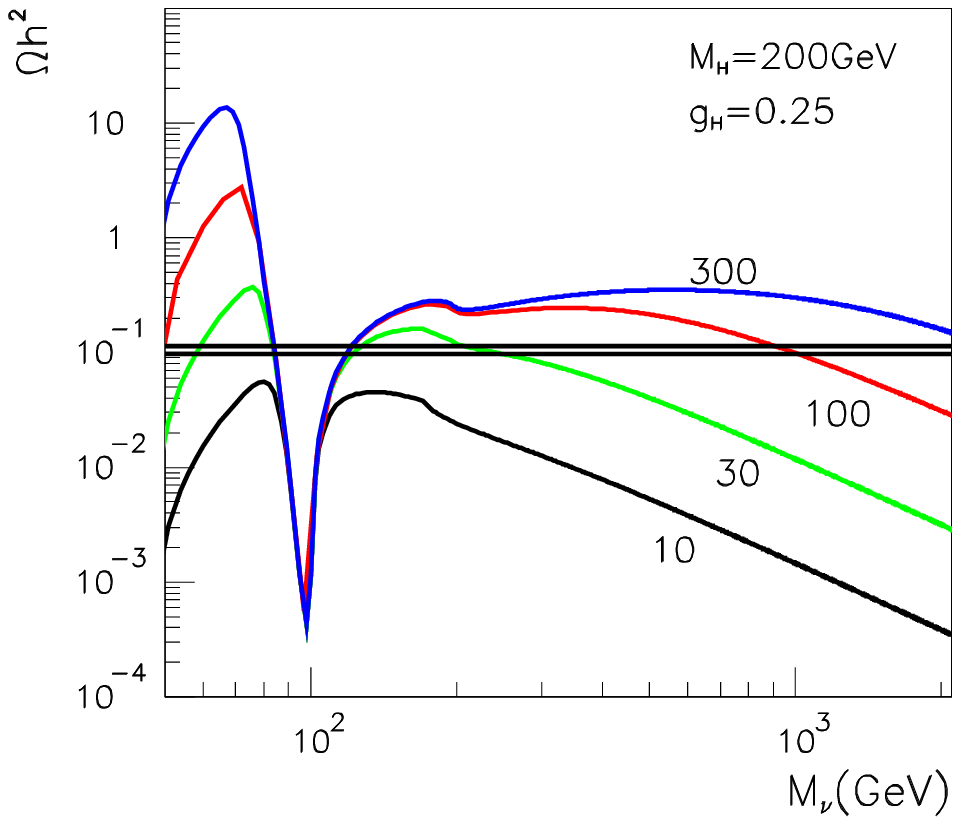,width=9.5cm,height=9.5cm}}
   \caption{ $\Omega_{\nup} h^2$ versus $\nup$ for
  $g/g_z=10,30,100,300,1000$ where  $g=e/(\sin \theta_W)$ is the SM coupling. In a), only the $ Z$-exchange is  included. For comparison, we also show in dotted line the relic density of a fourth generation Dirac neutrino with SM coupling to the $Z$ and a Yukawa coupling to the Higgs. For this curve, we set $m_H=200$ GeV. In b) the Higgs exchange is included as well with $g_H$=0.25. }
   \label{fig:gzcontours}
\end{figure}

\begin{figure}[!htb]
 \centerline{\epsfig{file=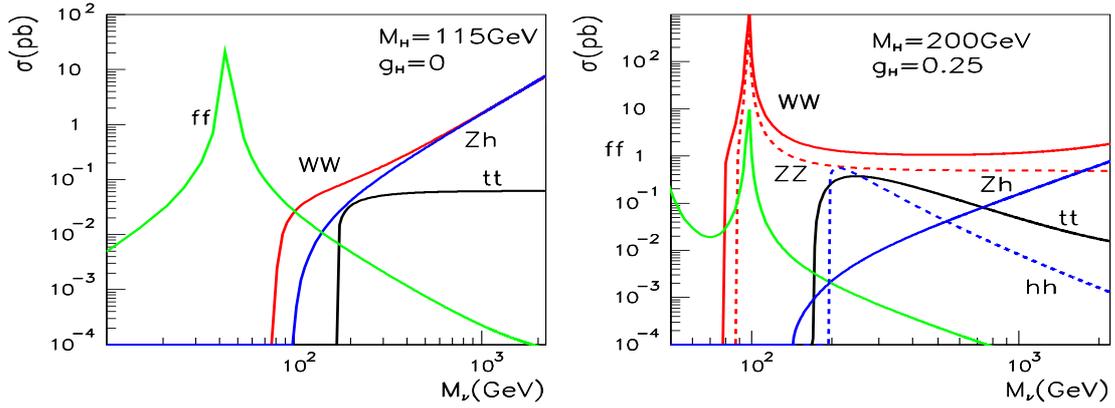, width=16.5cm,height=7.5cm}}
  \caption{Annihilation cross sections at freese-out.  The effect of a $Z'$ is omitted here and will be shown in Fig.~\ref{fig:cross}.} \label{fig:crosssectionsfreeseout}
\end{figure}

Annihilation channels of $\nup$ are listed in Fig.~\ref{fig:Feyn_Diag}.
We first look at the effect of the coupling to the $Z$, $g_Z$, and set $g_{Z'}=0$. For $\mnup \lsim 100 \mbox{ GeV}$, annihilation into fermions dominate due to the $Z$ resonance. For $\mnup \gsim 100 \mbox{ GeV}$, 
main annihilation channels are into $WW$ and $Zh$ (the relative contribution of
$Zh$ increases for larger $\mnup$). The contribution from top pairs is small.
Figure~\ref{fig:gzcontours}a shows how $\Omega_{\nup} h^2$ decreases as function of $\mnup$ for 
different $g_Z$ couplings.   In Fig.~\ref{fig:gzcontours}a, we show for comparison the prediction  for a fourth generation Dirac neutrino with SM coupling to the $Z$ and a Yukawa coupling to the Higgs as studied in \cite{Enqvist:1988we}.
Since $\nup$ has non-standard couplings, the total annihilation cross section grows with $\mnup$\footnote{This is different from the behaviour of the vector-like Kaluza-Klein neutrino studied in \cite{Servant:2002aq}.}. 
Unitarity breaks down for $\mnup$ at the multi TeV scale and we do not show any predictions beyond these values.
Figure~\ref{fig:gzcontours}b shows the effect of a coupling of $\nup$ with the Higgs\footnote{We neglect the contribution to the mass of $\nup$ due to the Yukawa coupling.}, in particular the resonance at $\mnup\sim M_H/2$. The contributions of the different annihilation channels are also displayed in 
Fig.~\ref{fig:crosssectionsfreeseout}. The $ZZ$ channel is important if  $\nup$ has a sizable coupling to the Higgs.
\begin{figure}[!htb]
  \centerline{\epsfig{file=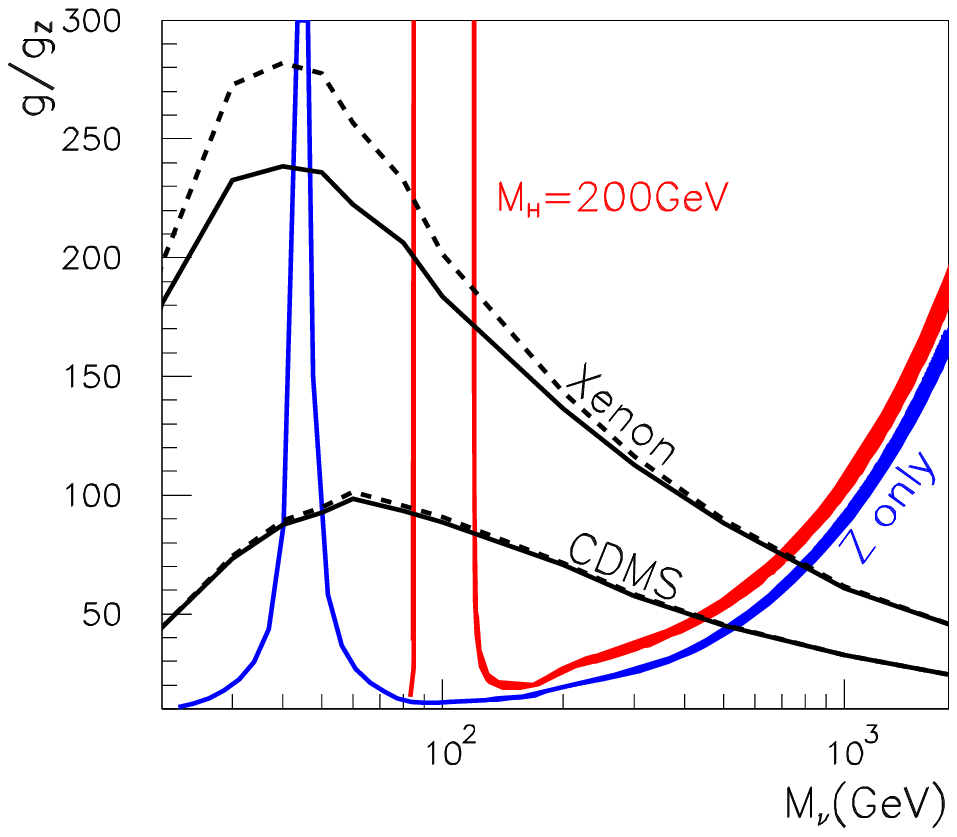, width=11cm,height=9.5cm}}
   \caption{WMAP region, $0.097<\Omega h^2< 0.113$, 
 in the $(\mnup,g/g_Z)$ plane (where  $g=e/\sin \theta_W$ is the SM coupling). The blue band includes the $Z$-exchange only, the red band includes the Higgs exchange as well. The  region excluded by CDMS and Xenon is also displayed. The dotted line takes into account the Higgs contribution.
 The region
  above the direct detection line and below the WMAP band is allowed. 
  In this figure, the effect of the $Z'$ is omitted. 
 } \label{fig:bilan}
\end{figure}

Near the $Z$ resonance, only a weak coupling is necessary
to get $\Omega_{\nup} h^2 \approx 0.1$. We combine the relic density constraints with the direct detection constraints in Figure~\ref{fig:bilan} where we show the 
WMAP \cite{Spergel:2006hy} allowed region 
($0.097<\Omega h^2< 0.113$)\footnote{A more conservative upper bound, $0.097<\Omega h^2<$0.134, was recently advocated in
\cite{Harmann}.} in the $g/g_Z-\mnup$ plane, as well as 
the CDMS and Xenon limits. The region satisfying both constraints corresponds to either
 $\mnup\approx$ 40--50 GeV or $\mnup \gsim 500$ GeV if the Higgs coupling is negligible.   For $\mnup
\gsim500$ GeV, the relic density
constraint on $g_Z$ is more severe than the direct detection constraint. If the Higgs coupling to $\nup$ is sizable, a region around $\mnup \sim M_H/2$ opens up. 

\subsection{Light neutrino scenario}
\label{subsec:lightneutrino}

\begin{figure}[!htb]
  \centerline{\epsfig{file=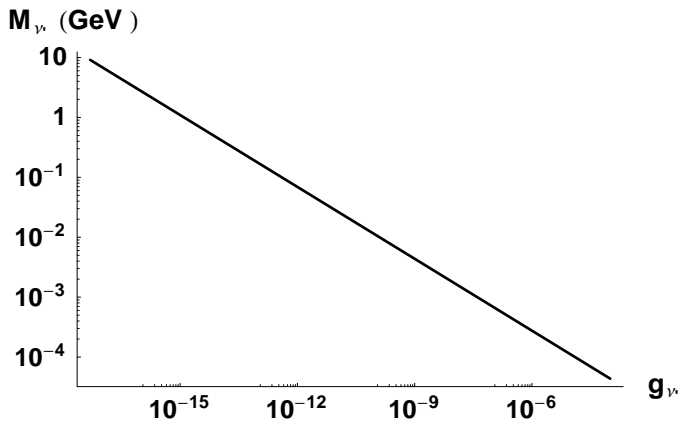, width=9cm}}
    \caption{$g_{\nup}$ is the typical size for the couplings  $Z\nup  \nu$ and  $ W \nup l^{\pm}$. Below the line, $\nup$ is cosmologically stable.
 } \label{fig:stableneutrino}
\end{figure}

For WIMP masses below 10 GeV, direct detection constraints do not apply. We concentrate on this particular case in this  section. Light WIMPs also potentially offer the interesting prospect that they can be stable without the need to introduce an extra symmetry, just because there is no SM particle they can decay into, at least on a cosmological time scale. In the absence of any particular discrete symmetry, our exotic light neutrino can have a three-body decay, like the muon. For instance, if the underlying theory is 5-dimensional with $SU(2)_R$ gauge symmetry, there are gauge couplings between the KK mode and the zero mode of $\nu_R$: $\nu_R^{(1)}Z'\nu_R^{(0)}$, $\nu_R^{(1)}W_R l_R^{(0)}$. They induce, via $Z-Z'$ and $W_R-W$ mixings, the effective 
couplings ($\nup$ then corresponds to $\nu_R^{(1)}$): $Z \nup  \nu^0_R$ and  $W\nup l^{\pm}$. If these are sufficiently suppressed, with strength of order $g_{\nup}$, the lifetime of $\nup $ can indeed exceed the age of the universe, as illustrated in Fig.~\ref{fig:stableneutrino}, using the width $\Gamma\sim (g_{\nup}/g)^2 G_F^2\mnup^5/(192\pi^3)$.
For a neutrino mass in the 1--10 GeV range, these coupling should be smaller than $10^{-15}$ for the neutrino to be cosmologically stable. This is actually naturally realized in Randall-Sundrum models where the zero mode  neutrino is localized near the Planck brane while the KK $\nu_R$ is peaked on the TeV brane. The overlap of their wave functions is therefore very suppressed, resulting in a tiny effective 4D coupling. However, in this case, we expect the KK $\nu_R$ to have a mass in the TeV range rather than 10 GeV. The only KK fermions that can be naturally light are those  belonging to the multiplet containing the top quark, as discussed in Part II. For instance, in warped GUT models, the KK $\nu_R$ belonging to  the {\bf 10} of $SO(10)$ which contains the top quark (that is the only zero-mode SM particle in the {\bf 10}) does not couple directly to any light SM fermion. The only SM fermion it can directly couple to is the top, via an $SO(10)$ gauge boson. Therefore,  its decay has to go through a very large number of intermediate states and will be very suppressed. In other words, the boundary conditions on the different components of the multiplet containing $\nu_R^{(1)}$ are such that no three-body decay is allowed.

Now the question is whether such a neutrino can naturally inherit the correct abundance.
The interesting aspect here is that these models typically offer the possibility that  the multi-body decay is suppressed while the self-annihilation can be large. This can be explained in terms of the different localizations for the wave functions of SM light fermions on one side and KK modes (of both fermions and gauge bosons) on the other side. 

It is clear from Fig.~\ref{fig:gzcontours}a that for $M_{\nup}\sim {\cal O}$(GeV),  $g_Z\sim {\cal{O}}(1)$ is needed  to obtain the correct thermal relic density; this is in contradiction with the experimental constraint from the invisible decay width of the $Z$ which requires $g_Z\lsim 0.035$. One way to open the window could be to consider that $\nup$ couples to a light ($\cal{O}$(GeV)) singlet scalar field $\phi$ that decays into SM fermions via its mixing with the Higgs. Near the resonance, $m_{\nup}\sim m_{\phi}/2$,  the right amount of annihilation can be obtained.
Another possibility is to assume that the reheat temperature is below the freese-out temperature $T_F\approx M_{\nup}/25$. For $M_{\nup}\sim 10$ GeV, $T_F\sim 0.4$ GeV, this is still well above the BBN temperature. In this case, the neutrino is produced through scattering in the plasma and the correct relic abundance may be reproduced with an appropriate choice of reheat temperature and $g_Z$ coupling. 
Note that in addition to $\nup$ production via the $g_Z$, $g_{Z'}$ and $g_H$ couplings, there is production of $\nup$ via the $g_{\nup}$ couplings defined above, which is similar to production via neutrino oscillation 
\cite{Dodelson:1993je,Dolgov:2000ew} but it is very suppressed given the $g_{\nup}$ values that we consider for cosmological stability.

Even if the lifetime of the neutrino exceeds the age of the universe, there are additional bounds on $g_{\nup}$ from EGRET and COMPTEL measurements of the diffuse gamma ray spectrum, that constrain the radiative decay with gamma emission $\nup\rightarrow \nu \gamma$  \cite{Kribs:1996ac}.
In conclusion, this $\cal{O}$(GeV) Dirac neutrino scenario would deserve a detailed analysis.

\subsection{Annihilation via t-channel heavy lepton exchange}
\label{subsection:leptoneffect}

\begin{figure}[!htb]
  \centerline{\epsfig{file=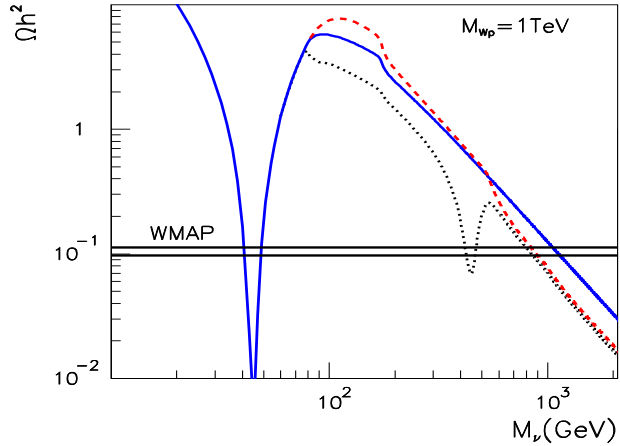, width=10cm,height=8.5cm}}
  \caption{ Effect of  the t-channel annihilation on $\Omega_{\nup} h^2$ for
  $g/g_z=100$. The blue full curve includes the $Z$ s-exchange only, the dashed (dot) curve
  includes the  t-channel  with  $M_{W'}=1$~TeV, $x_W=10$, leading to $g_W=0.065 $ and
  $M_{\tau'}/M_{\nup}=1.2$ without (with) coannihilations.
 } \label{fig:gw}
\end{figure}

We now consider  the effect of the t-channel exchange of a
heavy charged lepton $\tau'$ (potentially a Kaluza-Klein lepton, see also the toy model of Ref.~\cite{Schuster:2005ck}) leading to the annihilation into $W$. The coupling is defined as
\begin{equation}
\frac{g_W}{2} \bar\nup \gamma_\mu (1\pm\gamma_5)\tau' W^\mu \;\;\; , \ \ 
g_W=x_W \left( \frac{M_W}{M_{W'}} \right)^2= g_{W'} \times \theta
\end{equation}
where  $g_W$ is written here in terms of the $W-W'$ mixing resulting from EW symmetry breaking.
According to Fig.~\ref{fig:gw}, quite a large $g_W$ coupling is needed to
see the effect of these diagrams. For instance, if $g_W$ arises from $W-W'$ mixing and $M_{W'}=1$ TeV, then we need $x_W\gsim10$. We can make the
following observations:
\begin{itemize}
\item{} There is a destructive interference between the s and t channels leading to an increase in $\Omega_{\nup}
h^2$ compared to the case with $Z$ exchange only.
\item{} When one $W'$ can be produced in association with a $W$ in the annihilation, there is a 
decrease in $\Omega_{\nup} h^2$ (see the kink around $\mnup\sim$500 GeV).
\item{} When ${\tau'}$ is not too much heavier than $\nup$,
coannihilation effects can reduce the relic density. For illustration,
we show the case $M_{\tau'}/\mnup=1.2$ where  coannihilations more than
compensate for the increase of $\Omega_{\nup} h^2$ due to the t-channel exchange. In addition,
near the $W'$ resonance, there is a sharp drop in the relic density.
\end{itemize}

\subsection{Constraints on the $W'$}
If the $W'$ comes from a Left-Right model,
there are strong limits on the $W'$ mass and mixing  assuming a manifest LR symmetry i.e. the same mixing matrix in L
and R quark sectors. Typical limits on the mixing angle are
approximately $\theta<10^{-3}$ (assuming $g_{W'}=g$), while limits on the mass are around 1
TeV. The best limit is from $K_L-K_S$ mixing, $M_{W'}>1.6$~TeV \cite{Beall:1981ze}.
There are also Tevatron limits using the decay channel into
electron and right-handed neutrino leading to $M_{W'}\gsim 750$ GeV \cite{Abachi:1995yi}. In our analysis,
 we consider that only the coupling of $W'$ to the third generation is non-suppressed, as well-motivated in Randall-Sundrum models. In this case, the Tevatron constraints are weakened if we restrict the decay to the $\tau$
channel and indirect limits from $K$ or $\mu$ decays do not apply.
Nevertheless, if $W'$ couples to quarks of the third generation (as in Randall-Sundrum models) there is an important constraint from
$b\rightarrow s \gamma$ \cite{Cho:1993zb,Babu:1993hx}. The $W'$ leads to 
an enhancement of the $b\rightarrow s \gamma$ amplitude by a factor $m_t/m_b$.
Irrespective of the $W'$ mass, a limit on the mixing is $-0.015<\theta \times{g_{W'}/g}<0.003$ \cite{Cho:1993zb,Babu:1993hx}.
This assumes similar quark mixing matrices in the L and R sectors.
If the $W'$ coupling to leptons is similar to that of quarks,
the above constraint is incompatible with the values chosen in Fig.~\ref{fig:gw},  $x_W=10$, $M_{W'}=$ 1 TeV, corresponding to $g_{W'}=0.065$.
 Relaxing the assumption that the mixing matrix is the
same in the left and right sectors will not help sufficiently. 
 In conclusion, the $W'$ effects can be ignored.
 
\subsection{Annihilation via s-channel $Z$ and $Z'$ exchange}

\begin{figure}[!htb]
  \centerline{\epsfig{file=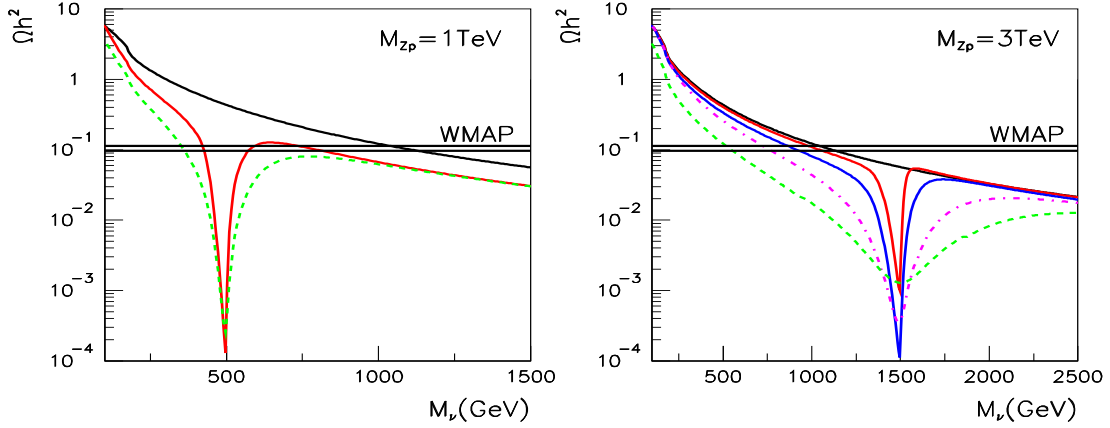, height=8.cm,width=16cm}}
  \caption{ $\Omega_{\nup} h^2$  for $g/g_Z=100$, $g_H=0$ and  $g_{Z'}=0$ (black), $g_{Z'}=0.3$ (red). Also shown  is the case where $Z'$ couples in addition to all third generation fermions (green dash) with $g_{Z'ff}=g_{Z'}=0.3$ in a) and  $g_{Z'ff}=1$ in  b). We also show in b) the case with $g_{Z'}=1$ (blue) and the case where $Z'$ couples to top quarks  with $g_{Z'tt}=1$ (pink/dash-dot).   Note that we fixed $\theta_{ZZ'}=10^{-3}$ in both a) and b) even though it is expected to scale as $M_Z^2/M_{Z'}^2$. 
 } \label{fig:Zp}
\end{figure}

We now consider the combined effect of the two annihilation channels through $Z$ and $Z'$.
We first look at the $Z'$ coupling to $W$ pairs, arising from $Z$-$Z'$ mixing:
\begin{equation}
g_{Z'WW}=g_{ZWW} \ \theta_{ZZ'}
\end{equation}
where $g_{ZWW} $ is the Standard Model coupling. 
Next we add the interaction of the $Z'$ with top quarks. There can be a noticeable reduction of the
relic density  due to this extra channel. Figure~\ref{fig:Zp} shows the effect of the $Z'$ exchange on $\Omega_{\nup} h^2$ for different values of the parameters. It opens a region at the $Z'$ resonance, for $\mnup\sim M_{Z'}/2$, and also reduces the upper bound on $\mnup$ from WMAP compared to the case with $Z$ exchange only.
\begin{figure}[!htb]
\centerline{\epsfig{file=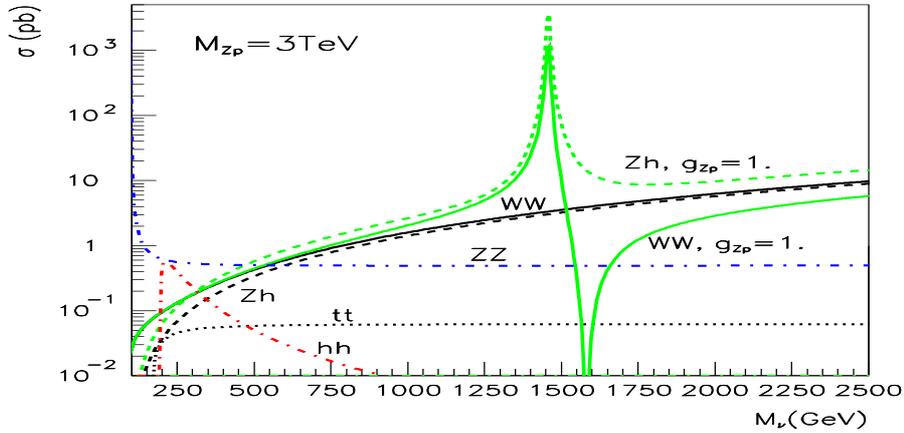, height=8.cm,width=14cm}}
\caption{ Annihilation cross-sections at freeze-out 
into $WW$ (solid), $Zh$ (dash), $ZZ$ (blue dash-dot), $hh$  (red dash-dot) and $t\bar{t}$ (dot) for  $g_Z=g/100$, $g_H=0.25$, $m_H=200$ GeV.  In black, only  the $Z$ exchange is included ($g_{Z'}=0$), in green, the $Z'$ exchange is included with $g_{Z'}=1$, $g_{Z'WW}=10^{-3}g_{ZWW}$ and $g_{ZZ'H}=0.5 g_{Z'} v$.
 } \label{fig:cross}
\end{figure}
We show in Fig.~\ref{fig:cross} the contributions of the different
channels to the total annihilation cross section. There is a destructive interference
between the $Z$ and $Z'$ contributions in the
$WW$ channel beyond the
$Z'$ resonance. However, this does not produce a significant  increase in the relic density, since the $Z'$ annihilation into $Zh$ is still important. In Fig.~\ref{fig:cross} we chose $g_{ZZ'H}=0.5 g_{Z'}  v$.

\subsection{Constraints on the $Z'$}
The coupling of  $\nup$ to the $Z$ can be induced, for example,  via $Z$-$Z'$ mixing or  via mixing with another heavy neutrino which has a large coupling to the $Z$.
If the $Z$-$Z'$ mixing comes from the Higgs {\it vev} only, then 
\begin{equation}
\theta_{ZZ'}= \frac{g}{2\cos \theta_W} \ g^H_{Z'} \ \frac{v^2}{M^2_{Z'}}
\label{theta_equation}
\end{equation}
In the absence of a custodial symmetry protecting the $\rho$ (or $T$) parameter, we have to impose the constraint
$\alpha T \approx \theta^2_{ZZ'} M^2_{Z'}/M^2_Z \lsim 10^{-3}$. When the $g_Z$ coupling is induced only via $Z-Z'$ mixing, then 
$\theta_{ZZ'}={g_Z}/{g_{Z'}}$ and it is difficult to satisfy the constraint on the $T$ parameter if $g^H_{Z'}\sim g_{Z'}$. 
A similar LEP constraint comes from the shift in the vectorial
coupling of the $Z$ to the $\tau$ and the $b$, leading to $\theta_{ZZ'} \ g_{Z'ff}< 4\ 10^{-3}$.  The $T$ parameter constraint  can be relaxed  in models with $SU(2)_R$ gauge symmetry which  are in any case a strong motivation for Dirac neutrino dark matter.  The constraint on the $Zbb$ coupling can also be evaded (see e.g. \cite{Agashe:2006at}).

Like the limits on $W'$,
most direct collider searches involve fermions of the
first and second generations  and if we assume that $Z'$ couples to the third generation only, we can tolerate $Z'$ as light as $\sim 500$ GeV \cite{Yao:2006px}.
 A $Z'$ which has generation-dependent couplings (like in Randall-Sundrum models) will induce tree-level FCNC. 
 If it couples to the third
generation only:
\begin{equation}
{\cal L} =( g_{Z'b_L} \bar{b} \gamma_\mu P_L b + g_{Z'b_R} \bar{b}
\gamma_\mu P_R b) Z'
\end{equation}
flavour non-diagonal couplings to the down-type quarks are induced
\begin{equation}
{\cal L} =( g_{Z'b_L} \bar{d_i} \gamma_\mu P_L  U^*_{L3i} U^*_{L3j}
d_j + g_{Z'b_R} \bar{d_i} \gamma_\mu P_R  U^*_{R3i} U^*_{R3j} d_j )
Z'
\end{equation}
with $U_{L,R}$ the mixing matrix for the left-(right)-handed
down-type quarks. This is turn will induce a flavour non-diagonal
coupling to the $Z$ due to $Z-Z'$ mixing. FCNC effects due to a $Z'$ with non-universal couplings
are analysed in Ref.~\cite{Langacker:2000ju,He:2004it}.
Constraints are model-dependent and can be avoided. They are typically weaker than the $T$-parameter constraint.
\begin{figure}[!htb]
  \centerline{\epsfig{file=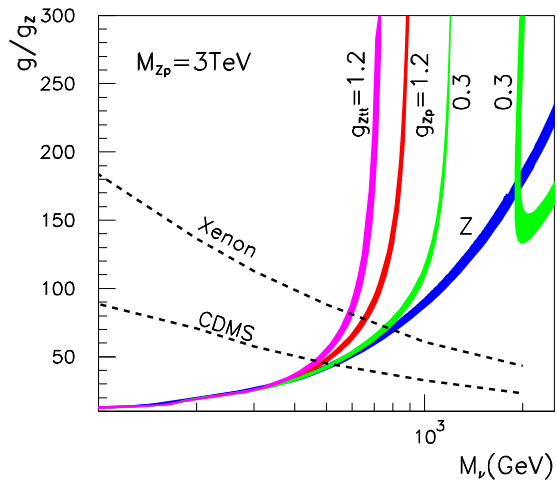, width=9.cm}}
  \caption{ WMAP region in the $(\mnup,g/g_Z)$ plane with $Z$ exchange only (blue), with $Z'$ echange for $g_{Z'}=0.3$ (green)  and $g_{Z'}=1.2$ (red).
The effect of adding a $Z'\bar{t}t$ coupling is also shown (pink) assuming $g^R_{Z'tt}=g_{Z'}=1.2$. In these plots, $\theta_{ZZ'}=10^{-3}$ and $g_H=0$.
The allowed region is above the XENON dashed line and below the  $\Omega h^2$ bands. }
  \label{fig:gzzp}
\end{figure}
\subsection{Contour plots}

\begin{figure}[!htb]
  \centerline{\epsfig{file=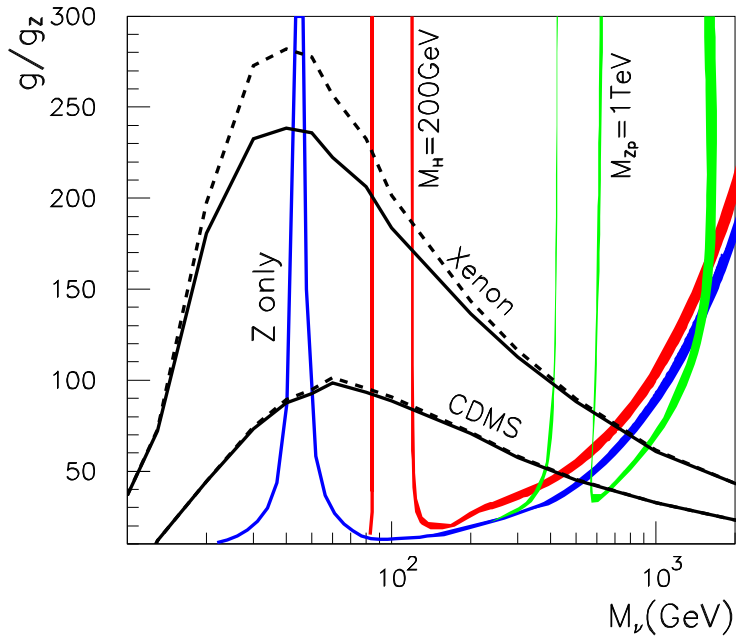, width=12.cm}}
  \caption{ Summary plot: WMAP region in the $(\mnup,g/g_Z)$ plane with $Z$ exchange only (blue), with $Z'$ echange for $g_{Z'}=0.3$ and $M_{Z'}=1$ TeV (green)  and Higgs exchange with $g_{H}=0.5$ (red).
In this plot, $\theta_{ZZ'}=10^{-3}$. The CDMS and XENON direct detection constraints are shown with (dashed) and without (solid) a $\nup$-Higgs coupling.
The allowed region of parameter space is above the XENON line and below the  $\Omega h^2$ bands.}
  \label{fig:plot_bilan}
\end{figure}

\begin{figure}[!htb]
\centerline{\epsfig{file=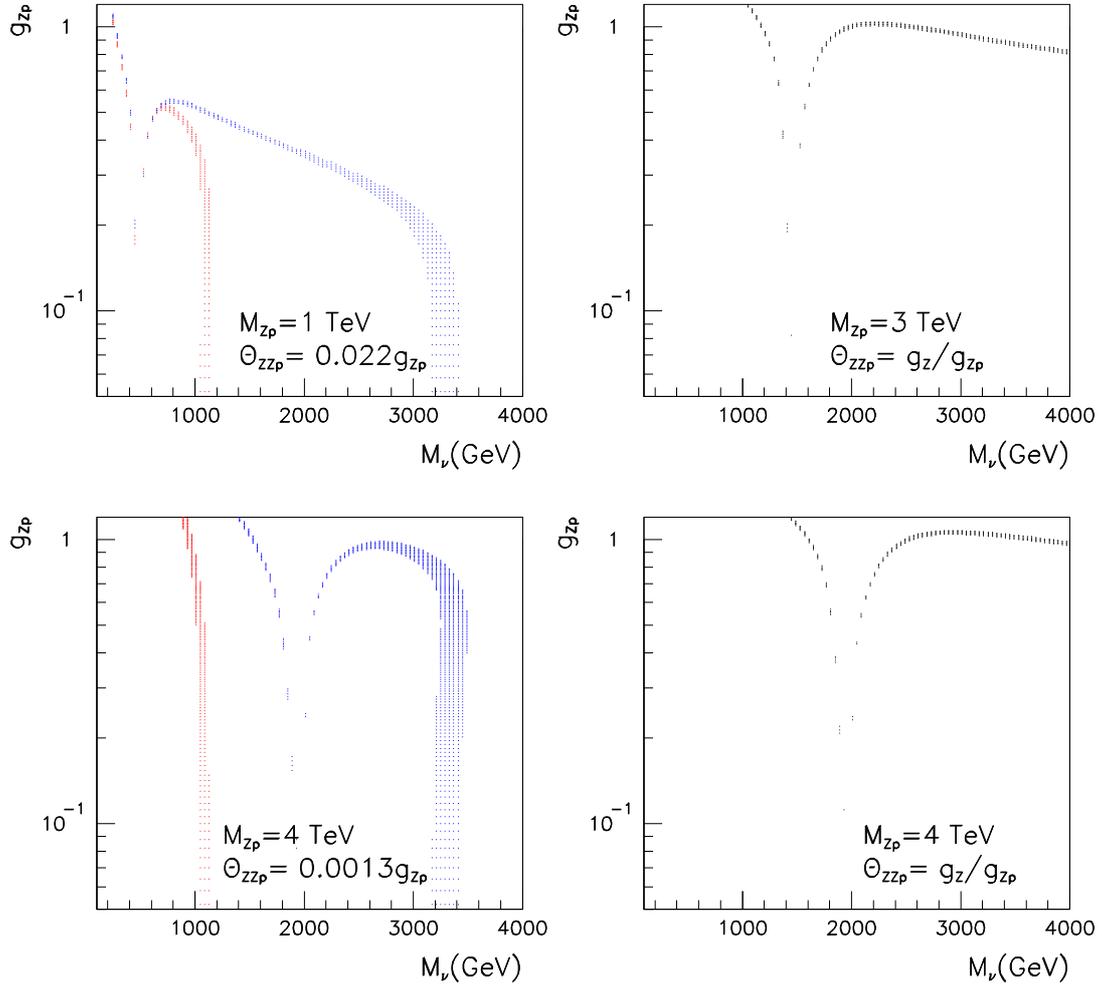, width=16cm}} \caption{WMAP region in the  $(\mnup,g_{Z'})$ plane. In all cases, $\theta_{ZZ'}=0.36 g^H_{Z'} (v/M_{Z'})^2$ and $g^H_{Z'}=g_{Z'}$.
In the two plots on the left, $g_Z$ is fixed to 
 $g_Z=g/100$ (red) and $g_Z=g/300$ (blue) independently of the $g_{Z'}$ value. In the two plots on the right,  $g_Z$ is fully due to $Z$-$Z'$ mixing.
 } \label{fig:gzmlzp}
\end{figure}

\begin{figure}[!htb]
\centerline{\epsfig{file=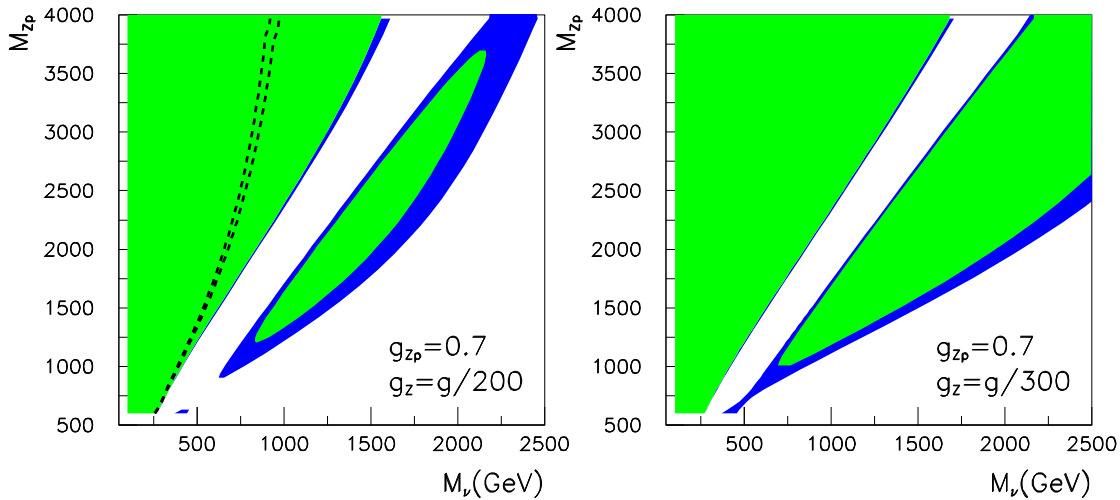, width=16cm}} \caption{
Relic density in the $(\mnup,M_{Z'})$  plane for $\mixzzp=10^{-3}$, $g_H=g_{Z'}=0.7$, $m_H=115$ GeV and we neglect the coupling of the $Z'$ to fermions. The blue (dark grey band) corresponds to the WMAP range
while in the green (light grey band) $\Omega_{\nup} h^2>0.113$ and in the white region  
$\Omega_{\nup} h^2<0.097$. On the first plot,  the contours of $\Omega_{\nup} h^2=0.097,0.113$ 
for  $g_Z=g/100$ and  $\mixzzp=g_z/g_{z'}$ (dash) are also shown. 
 } \label{fig:mzpmlzp}
\end{figure}

We summarize our results and show the WMAP region in three different planes: $(\mnup,g/g_Z)$ in Fig.~\ref{fig:gzzp} and \ref{fig:plot_bilan},  $(\mnup,g_{Z'})$ in Fig.~\ref{fig:gzmlzp} and $(\mnup,M_{Z'})$ in Fig.~\ref{fig:mzpmlzp}.
In Fig.~\ref{fig:gzmlzp} and ~\ref{fig:mzpmlzp}, we neglect for simplicity the coupling
of $Z'$ to the fermions. 
In  Fig.~\ref{fig:gzzp} and \ref{fig:plot_bilan}, we fixed $\theta_{ZZ'}=10^{-3}$ independently of the relation 
(\ref{theta_equation}) while the two plots  of Fig.~\ref{fig:gzmlzp} satisfy Eq. (\ref{theta_equation}).
 In addition, the two right-handed plots of Fig.~\ref{fig:gzmlzp} assume that the only source of 
the $g_Z$ coupling is the $\theta_{ZZ'}$ mixing. 
 As a result, the $\nup-Z$ coupling is suppressed and $\Omega_\nup h^2$ falls within 
the WMAP range only  for a large  coupling to $Z'$, $g_{Z'}\approx 1$, or for $\mnup \approx M_{Z'}/2$.
The situation is best summarized in
Figure \ref{fig:plot_bilan} which captures what is the allowed region of parameter space after imposing the WMAP bound and XENON constraints: There is a small mass window  allowed for $\mnup\approx M_Z/2$. There are other wider mass windows  near $M_H/2$ and $M_{Z'}/2$. Away from these resonance effects, a large region opens up for $\mnup \gsim 700 \mbox{ GeV}$.
\newpage
\section{Collider signatures}

Like in other WIMP models, the standard searches rely on pair production of the heavier exotic particles which ultimately decay into the WIMP, leading to signals with energetic leptons and/or jets and missing $E_T$.
We list below some signatures which are more specific to the neutrino WIMP model. 
\subsection{Invisible Higgs decay into $\nup$}
As seen previously, if $\nup$ has a significant coupling $g_H$ to the Higgs, it  can account for the dark matter of the universe for $m_{\nup}\sim m_H/2$ corresponding to the Higgs resonance. A  significant  $g_H$ coupling can arise for instance in the model of \cite{Agashe:2004ci,Agashe:2004bm} (see section 9.3).

\begin{figure}[!htb] 
  \centerline{\epsfig{file=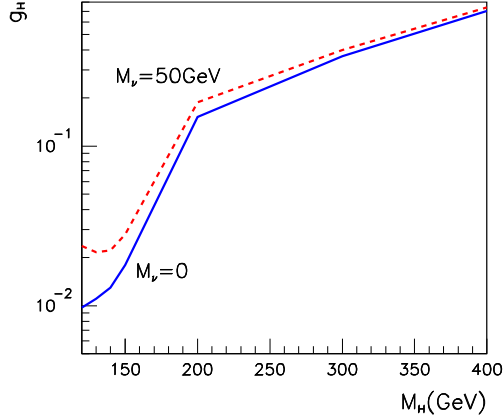, width=8cm}}
  \caption{Above the lines, 
  an invisible Higgs would be probed at LHC with ${\cal L}=100 \mbox{ fb}^{-1}$.
 } \label{fig:hinv}
\end{figure}

As a result, the Higgs can decay invisibly into $\nup$ with a significant branching fraction.
 The $g_H$ coupling could be probed at the LHC in Higgs production associated
 with gauge bosons  \cite{Godbole:2003it}, or with top quarks  
or in the  weak boson fusion process \cite{Eboli:2000ze}. 
The weak boson fusion process seems to be the most promising. In the following we use the results of Ref.~\cite{Cavalli:2002vs} where
an analysis including a  detector simulation was performed and a
limit on the invisible width was obtained for various Higgs masses. In general,  this limit
takes into account the fact that   the production cross-section could be modified
relative to the SM one,  here we assume that  the couplings of the Higgs to quarks
are the standard ones.
A more recent study \cite{Davoudiasl:2004aj} performed a 
refined analysis of the $ZH$ channel, combined with the boson
fusion channel. For  light Higgses ($m_H<160$GeV) the results are similar to the
ones of Ref.~\cite{Cavalli:2002vs}. 
The partial width of the Higgs into neutrinos is
\begin{equation}
\Gamma (h\rightarrow \nup\overline{\nup}) =  \frac{g^2_{H}}{8 \pi}M_H
\left(1-4\frac{M_{\nup}^2}{M_H^2} \right)^{3/2}
\end{equation}
Fig.~\ref{fig:hinv} shows the resulting limit that can be obtained on the $g_{H}$ coupling at LHC.
\begin{figure}[!htb] 
  \centerline{\epsfig{file=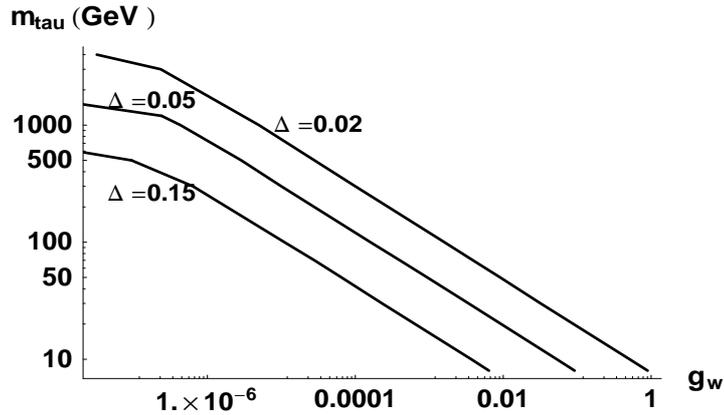,  height=5.5cm,width=9.5cm}}
  \caption{Lines delimiting the region where $\tau'$ is too long-lived to decay inside the detector for three values of  $\Delta$, the relative mass splitting between $\nup$ and $\tau'$.}
 \label{fig:taurcontours}
\end{figure}
\begin{figure}[!htb] 
  \centerline{\epsfig{file=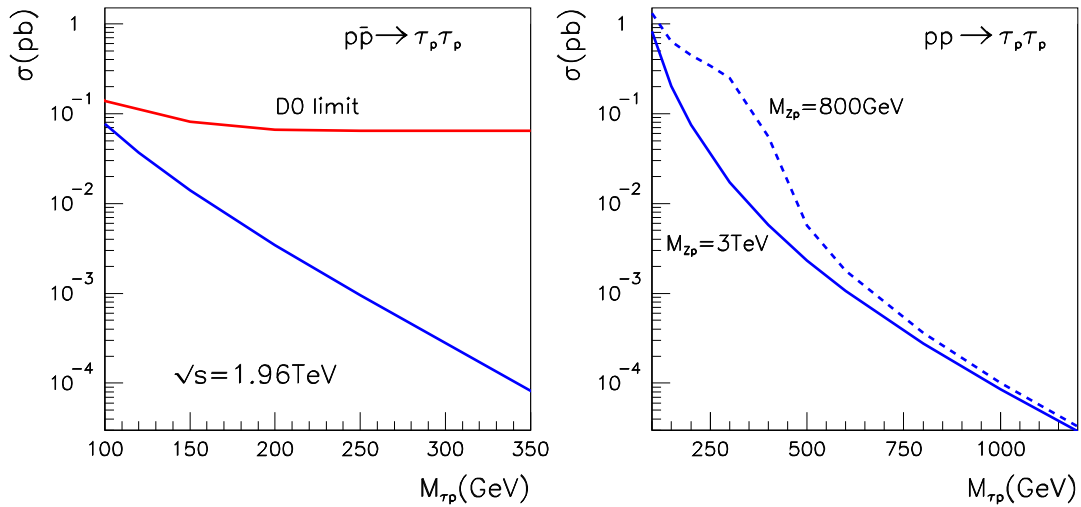, width=16cm}}
  \caption{ $ pp\rightarrow \tau^{\prime+}\tau^{\prime-}$ cross section at the Tevatron and the LHC. $M_{Z'}=3$ TeV (solid line) and $M_{Z'}=800$ GeV (dashed line). The coupling of $Z'$ to $b$ is $g_{Z'b}=1$.  }
 \label{fig:taurproduction}
\end{figure}
\subsection{$\tau'$ production}
Let us consider models where $\nup$ belongs to a gauge multiplet, for instance an $SU(2)_R$ multiplet. If $\nup$ and its $\tau'$ partner are nearly degenerate in mass, $\tau'$ is long-lived and 
the pair production of $\tau'$ could lead to interesting stable CHAMP\footnote{Charged Massive Particles}-like signatures. This is to be contrasted with the standard scenario in which the dark matter is dominantly produced through decays of colored particle and therefore accompanied by energetic jets rather than by charged tracks.
This situation was  also discussed in the dark matter model of \cite{Cirelli:2005uq}.
Fig.~\ref{fig:taurcontours} shows the region in the $(g_{W\tau'\nup},m_{\tau'})$ plane where $\tau'$ decays outside the detector.
Limits  from LEP \cite{Abbiendi:2003yd} ($m_{\tau'}\gsim 100$ GeV) and D0  \cite{D0champs} are reproduced on our plot (Fig.~\ref{fig:taurproduction})  of  the $\tau'$ pair production cross section, which is dominated by the $Z$ exchange since $Z'$ only couples to the third generation.
A more likely possibility is that $\tau'$ will decay inside the detector. In this case, the search is similar to that of sleptons with signature {\it two leptons} $+E_T^{\mbox{miss}}$ (see Ref.~\cite{Abbiendi:1999sa,Abbiendi:2003ji} for LEP constraints). 

\section{Conclusion of Part I}
In summary,  Dirac neutrinos  are viable dark matter candidates. 
The situation is best summarized in Figure \ref{fig:plot_bilan}.
For a  mass between 10 GeV and 500 GeV, the main requirement is that the coupling to the $Z$ should be at least 100 times smaller than the SM neutrino-$Z$ coupling  in order to satisfy the direct detection constraint. Once this is satisfied,  there is a large range of neutrino and $Z'$ masses as well as $Z'$ couplings that lead to the correct thermal abundance. The annihilation via $Z$  is the dominant mechanism
 for $\nup$ masses below 100 GeV. Near $M_Z/2$, the
annihilation mechanism is even too efficient. If $\nup$ has a large coupling to the Higgs, $\mnup\sim M_H/2$ can lead to the correct relic density. Finally,  $\nup-Z'$ couplings  open a large spectrum of $\mnup$ possibilities in the multi hundred GeV range up to the $Z'$ mass.

In this work, we have assumed that there is no primordial leptonic asymmetry in the dark matter sector ($n_{\nup}=\overline{n}_{\nup}$). Obviously, our predictions for the relic density could change significantly if there was such an asymmetry, like there is in the visible matter sector. In constrast, this issue does not arise  with neutralino dark matter or heavy Kaluza-Klein gauge boson dark matter.
We have also restricted our analysis to the case where Dirac neutrinos would constitute all the dark matter. Constraints would be relaxed if we assumed instead that they constitute only a subdominant piece of dark matter\footnote{The scenario where a fourth generation neutrino with mass near the $M_Z/2$ window is a subdominant component of dark matter was studied in \cite{Belotsky:2004st}.}.
Moreover, we only studied the case where Dirac neutrinos are thermal relics. Very different conclusions can be drawn if instead, the production mechanism is non thermal, and this is left for a future project.

Except in Section \ref{subsection:leptoneffect}, we have not considered coannihilation effects, as this is a more model-dependent issue. They will be studied in the explicit example of Part II.
Finally, we have not discussed indirect searches in this work, this was done in \cite{Hooper:2005fj} for the case of the LZP model that we now present in details. \\

\bigskip

{\Huge{\bf Part II - An explicit example: The LZP in warped GUTs}}\\
\label{section:LZP} 

In Ref.~\cite{Agashe:2004ci,Agashe:2004bm} it was shown that in models of warped extra-dimensions
embedded in a GUT, the symmetry introduced to prevent rapid proton
decay, a $Z_3$ symmetry, also guarantees the stability of a
light KK fermion, a KK right-handed neutrino.
This particle is called the LZP and its
 properties have been
studied in ~\cite{Agashe:2004ci,Agashe:2004bm}. A detailed analysis of the
indirect detection prospects in neutrino telescopes, cosmic positron experiments and gamma ray telescopes  was also presented in
~\cite{Hooper:2005fj}  and in \cite{Barrau:2005au}  for antiproton experiments. Some collider signatures were discussed in \cite{Agashe:2004bm} and more recently in \cite{Dennis:2007tv}.
In this paper, we revisit the properties of the LZP and perform a
complete calculation of its relic density. 

The underlying model is based on the Randall--Sundrum setup \cite{Randall:1999ee}, 
where the hierarchy between the
electroweak (EW) and the Planck scales arises from a warped higher dimensional
spacetime. All Standard Model (SM) fields except the Higgs  (to solve the
hierarchy problem, it is sufficient that just the Higgs --or alternative
dynamics responsible for electroweak symmetry breaking-- be localized at the
TeV brane) have been promoted to bulk fields rather than brane fields.  EW precision constraints require  the EW gauge
symmetry in the 5-dimensional bulk to be enlarged to $SU(2)_L \times SU(2)_R \times U(1)_{X}$
\cite{Agashe:2003zs}.  
The AdS/CFT correspondence suggests that this model is
dual to a strongly coupled CFT Higgs sector \cite{Arkani-Hamed:2000ds}. Also,
the  $SU(2)_L \times SU(2)_R$ gauge symmetry in the RS bulk implies the
presence of a global custodial isospin symmetry of the CFT Higgs sector, thus
protecting EW observables from excessive new contributions
\cite{Agashe:2003zs}. 

In this framework,  Kaluza-Klein (KK) excitations of gauge bosons of mass
$M_{KK} \sim 3 $~TeV are allowed and interestingly, light KK fermions are expected in the spectrum
 as a consequence of the
heaviness of the top quark. The heaviness of the top quark is explained by the localization of the wave function of the top quark zero mode near the TeV brane. This is done by choosing a small 5D bulk mass (the so-called ``c" parameter), guaranteeing a large Yukawa coupling with the Higgs.
It is clear that if we want to embed these models into a GUT, we cannot consider $SU(5)$  but rather   Pati-Salam or $SO(10)$ schemes.
When the Right-Handed (RH) top quark is included in a GUT multiplet, its KK partners do not have a zero mode but their first KK excitation turn out to be light.
The masses and some of the couplings of these KK fermions are determined by the
$c$-parameters (see Eq.~\ref{formula:mLZP} in the appendix). 
To have
${\cal O}(1)$ top Yukawa, the right-handed top  must have
$c_{t_R}<0$.  We fix the parameters associated with the
top quark to be $c_{t_L}=c_L=0.4,c_{t_R}=c_R=-0.5$.
Among all SM particles, the $c$ of the RH top quark is the smallest. As a consequence, the KK modes inside its multiplet are predicted to be light (see Eq.~\ref{formula:mLZP}). 
They are likely to be the lightest KK states in these models and could be produced at the LHC \cite{Agashe:2004bm,Dennis:2007tv}. 

The main feature of unification in extra dimensions is that the SM fermions have to be split into different GUT multiplets (unless the SM fermions are localized on the Planck brane). However, this is still not enough to prevent proton decay from dangerous higher dimensional operators localized on the TeV brane.  
In ~\cite{Agashe:2004ci,Agashe:2004bm}, the problem of baryon number violation was solved by imposing
a $Z_3$ symmetry. The consequence of the $Z_3$ symmetry is also to provide a stable particle.
In this model, for each fermion
generation,  there are at least  three {\bf 16} multiplets. Each of them is assigned a baryon number. For instance, for the third generation:
\begin{eqnarray}
B=&1/3& \;\;{\bf t_L, b_L},t^{c}_R,t^{c}_R,\nu_L,\tau_L,\tau^{c}_R,\nu^{c}_R\nonumber\\
B=&-1/3 & \;\;t_L, b_L,{\bf t^{c}_R,b^{c}_R},\nu_L,\tau_L,\tau^{c}_R,\nu^{c}_R\nonumber\\
B=&0& \;\;t_L, b_L,t^{c}_R,b^{c}_R,{\bf
\nu_L,\tau_L,\tau^{c}_R,\nu^{c}_R} \label{eq:particles}
\end{eqnarray}
where the particles in bold have zero modes and correspond
to the SM fermions. The $Z_3$ symmetry is
\begin {equation}
\Phi \rightarrow e^{2\pi i \left( B-\frac{n_c-n_{\bar
c}}{3}\right)}\Phi \label{eq:Z3}
\end{equation}
where $n_c$ ($n_{\bar{c}}$) is the number of color indices of the (anti)particle $\Phi$.
SM particles do not carry any $Z_3$ charge. The Lightest $Z_3$ charged Particle (LZP) is therefore stable.

Although there are many new fermions
in this model, we focus on the lightest ones, i.e. the level one KK fermions that
belong to the multiplet containing the SM $t_R$ (the $B=-1/3$
multiplet of the third generation in Eq.~\ref{eq:particles}). 
The LZP belongs to this multiplet. The only phenomenologically acceptable choice is that $\nur$ is the LZP ($\nu_L$ couples too strongly to the $Z$). 
All fermions inside a given multiplet should have the same $c$-parameter. However, because of
bulk GUT breaking effects, there can be effectively  large splittings so that the $c$-parameters of each component in the multiplet can be treated as free parameters. 
The $c$-parameters
of members of $SU(2)$ doublets are
identical, for example $c_{t_L}$ for both $t_L^{(1)},b_L^{(1)}$.
The model therefore includes six free parameters for the third
generation KK fermions,
\begin{equation}
\label{eq:c}
c_{\nu_R},c_{\tau_L},c_{\tau_R},c_{b_R},c_{t_L},c_{\nu'_L}
\end{equation}
We ignore  the KK partner of the right-handed top. $t_R$ has 
different boundary conditions to provide the zero mode of the RH top quark and its first KK mode
is in the multi-TeV range.

 We also consider the influence of
the $\nul$ of the $B=1/3$ multiplet which plays a role because of
the mixing induced to the LZP. We will not include this particle
in the model or compute explicitly its contribution to scattering
processes but we will take into account its mixing with the LZP
which will influence the $Z$-LZP coupling.

Among the 45  gauge bosons  of $SO(10)$, we only consider the
gauge bosons of $SU(2)_L\times SU(2)_R\times U(1)$,
$W^\pm_L,W^0_L,W^\pm_R,W^0_R,B'$ which neutral components
recombine to give $Z,Z',\gamma$, and a leptoquark gauge boson of
electric charge $Q=2/3$ which belongs to a color triplet + anti-triplet
invariant under $SU(2)$'s, $X_s,\bar{X}_s$. This state has $Z_3$
number $\pm1/3$. Other gauge bosons will not directly enter the
$2\rightarrow 2$ annihilation cross-sections that are relevant for us.
The masses of these KK gauge bosons are taken to be all equal,
$\mxs=\mzp=M_{W'}=M_{KK}$.

The parameters that we expect to have an effect on the
annihilation rates are:  $g_{10}$ , the 4D $SO(10)$ coupling which determines the strength of
the $Z'$ couplings (thus the coupling of the $Z$ to the LZP) and can be considered as a free parameter; $\cnur$
which determines the LZP mass;  $\cnul$ which enters the
$\tilde{\nu}'_L-\nu_R$ mixing thus affects the $Z$-LZP coupling; the Higgs
mass which is relevant mainly when $M_{LZP}\approx M_H/2$ and
$M_{KK}$ which sets the mass of the new gauge bosons, in particular $Z'$ and $X_s$. 

We have  in
addition, $r$, parametrizing the amount of bulk GUT breaking, and $\Lambda$ the cut-off scale of the effective 4D theory.  Both enter the expression of
 the $X-X_S$ mixing (see Eq.~\ref{thetaXmixing}) thus the LZP coupling to
$t_L X_S$. The mass of $W'$ should be important when it can
contribute to a coannihilation process near a resonance. Finally,
masses of KK fermions should be mostly relevant when they are near
$M_{LZP}$ and contribute to coannihilation.

All the formulae for the calculation of masses and couplings
needed for the implementation of the model into \calchep~\cite{Pukhov:2004ca} are listed
in the appendix.

\begin{figure}[!htb]
  \centerline{\epsfig{file=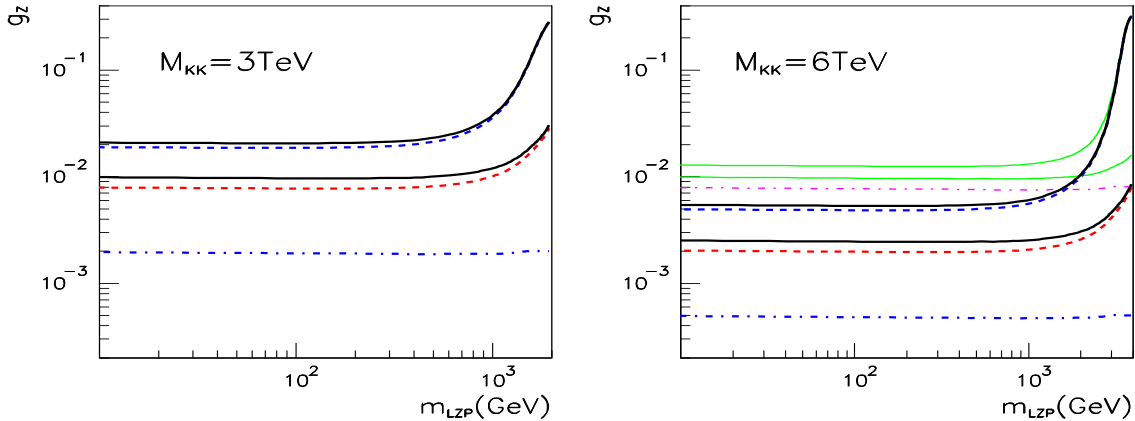, height=8cm,width=17cm}}
  \caption{ LZP-$Z$ coupling  as a function of the LZP mass.
   We show separately the contribution from $Z$-$Z'$ mixing (dash-dot) and 
$\nu_R$-$\nu'_L$ mixing (dash) as well as the total coupling $\gz$ (full).
 a) $M_{KK}=3$~TeV,  $g_{10}=0.3$; 
  b) $M_{KK}=6$~TeV, $g_{10}=0.3$ (blue/dark)
  and $g_{10}=1.2$ (green/light grey).
Lower (upper) curves are for $c_{\nu'_L}=0.4$ $(-0.1)$.
 } \label{fig:gz}
\end{figure}

The $Z$ coupling to the LZP is induced via $Z-Z'$
mixing as well as via the mixing with $\tilde{\nu}'_L$, the LH KK neutrino that belongs to the $B=1/3$ multiplet that contains the SM LH top.
The $Z'$
coupling to the LZP is of order 1 and proportional to
$g_{10}$. The $Z-Z'$ mixing therefore increases with $g_{10}$ and the resulting induced LZP-$Z$ coupling goes as $g_{10}^2$.
The  $\nul-\nu_R$ mixing
goes approximately like $\mlzp/\mnul^2$ thus is large when the
 the left-handed neutrino is light. This corresponds to
$\cnul=-0.1$. Furthermore, the mixing increases significantly when
the mass of the LZP approaches that of the $\nul$.
Figure ~\ref{fig:gz} shows how the two components of the  $Z$-LZP
coupling vary with $\mlzp$. The component which arises from
$\nu_L-\nu_R$ mixing clearly dominates, and  steeply increases when
$\mlzp >1 $~TeV.

The Higgs coupling to the LZP is also  suppressed by a
KK mass. Explicitly, in the limit $M_{\nu'_L}>>\mlzp$,
\begin{equation}
\label{formula:Higgscoupling}
g_H=
\frac{m_t  \ M_{LZP}}{M^2_{\nu'_L}(1-2\ccl)(1-2\ccr)} \times \sqrt{\frac{1-2\cnur}{2}}
\end{equation}
where $M_{\nu'_L}\approx M_{KK}$ and the term under the square root is there only  if $\cnur<-1/2$. Exact expressions used in the
code can be found in the Appendix. This coupling increases with
$\mlzp$ so  the largest contribution of the Higgs
exchange is expected for a heavy LZP. On the other hand,  the Higgs
contribution is  important for $\mlzp\approx M_H/2$.

\section{Relic density}

In this section, we compute the LZP relic density and explore the parameter space of the model.
The new features of our calculation  compared to Ref.~\cite{Agashe:2004ci,Agashe:2004bm}
are the following:
\begin{itemize}
\item{} We include all annihilation and coannihilation channels
involving  level 1 KK fermions of the third generation (with
the exception of the KK partner of the right-handed top that is heavy) that belong to the $t_R$ multiplet. We
include as well the exchange of level 1 KK gauge bosons,  $W',Z'$ and
vector leptoquarks, $X_S$.
\item{} We include all possible resonances in either annihilation and
coannihilation channels and perform precise evaluation of
annihilation cross-section near resonances, in particular the
Higgs resonance which was neglected in ~\cite{Agashe:2004ci,Agashe:2004bm}.
\item{} We exactly solve the Boltzmann equation for the LZP number density. Specifically, we do not rely on the non-relativistic approximation
$\sigma=a+b v^2$ (which fails near the resonances).
\end{itemize}
We achieve this using \micromegas. We
have rewritten the model in the
\calchep \ notation, specifying the new particles and their
interactions. All details can be found in the appendix. Since we
have a $Z_3$ symmetry rather than a R-parity,  special
care had to be taken to specify the particles that could
coannihilate with the LZP. We included all level one KK
fermions. The diagrams that contribute to each (co)-annihilation
process are chosen automatically according to the interactions
specified in the model file. We have also written a module for the
direct detection cross-section.
This only includes the dominant
contribution that arises from the $Z$-exchange diagram. 
The value of $g_H$, Eq.~(\ref{formula:Higgscoupling}), in the LZP model is indeed typically too small for the Higgs exchange to contribute, according to the analysis of Part I. 

 Unless otherwise noted, we consider the following range of values for the
free parameters of the model:\\
\begin{tabular}{lllll}
3 TeV$<M_{KK}<10$ TeV, & & $115<M_H<500$ GeV, && $2M_{KK}<\Lambda<6M_{KK}$\\
$-0.1<\cnul<0.9$, && $0<r<0.2$, && $0.3<g_{10}<1.2$\\
\end{tabular}\\
The central value
$\cnul=0.4$ is fixed by the zero mode of the LH top quark $c_L$ and
we allow deviations from bulk breaking effects ($\pm 0.5$).
The lower value for $M_{KK}$ is constrained by EW precision tests and direct detection
experiments. The upper value is set
arbitrarily to concentrate on models that are relevant to low energy
phenomenology.

A light LZP could contribute to the invisible width of
the $Z$. We have taken this constraint into account,
$\Gamma_Z^{inv}<1.5$ MeV, and found that it plays a role only for the 
near maximal value of $\gten\approx 1.2$, and $M_{KK}\approx
3$~TeV.

\subsection{LZP annihilation}

We first look at  self-annihilation  before analysing the
impact of various fermion coannihilation channels.  
%
%
\begin{figure}[!htb]
  \centerline{\epsfig{file=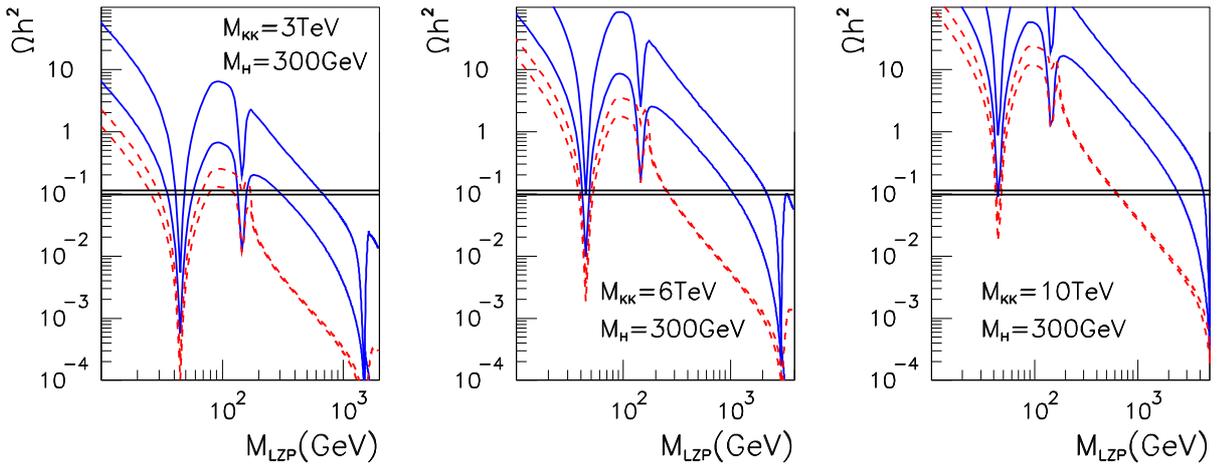, width=18cm}}
  \caption{Relic density  of the LZP without coannihilation. The blue full curves correspond to $g_{10}=0.3$
   and the red dashed curves to $g_{10}=1.2$. Lower (upper) curves are for $\cnul=0.9 (-0.1)$.
   The WMAP preferred region corresponds to the
  horizontal lines. Here,  the mass of all other KK fermions is $0.9\times M_{KK}$.} \label{fig:omlzp}
\end{figure}
\begin{figure}[!htb]
 \centerline{\epsfig{file=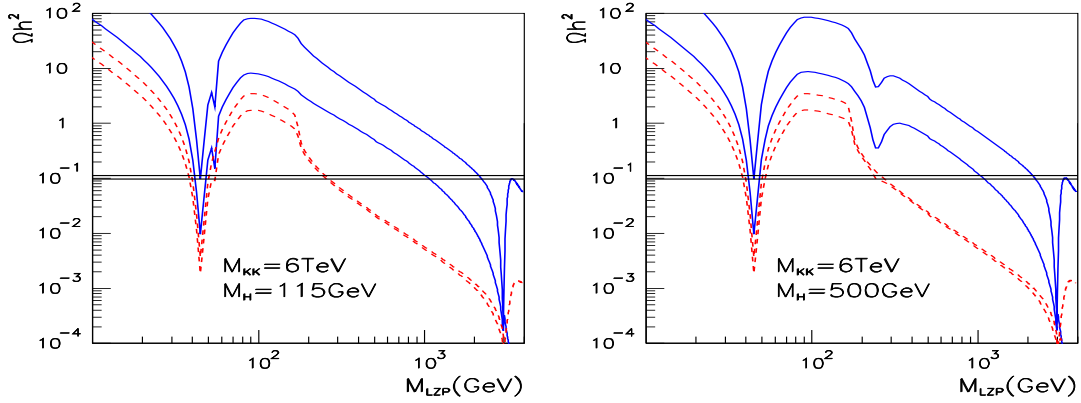, height=7cm,width=16cm}}
 \caption{Same as Fig.~\ref{fig:omlzp} with a) $M_H=115$~GeV, b) $M_H=500$~GeV.} \label{fig:omlzp6}
\end{figure}
%
%
%
\begin{figure}[!htb]
  \centerline{\epsfig{file=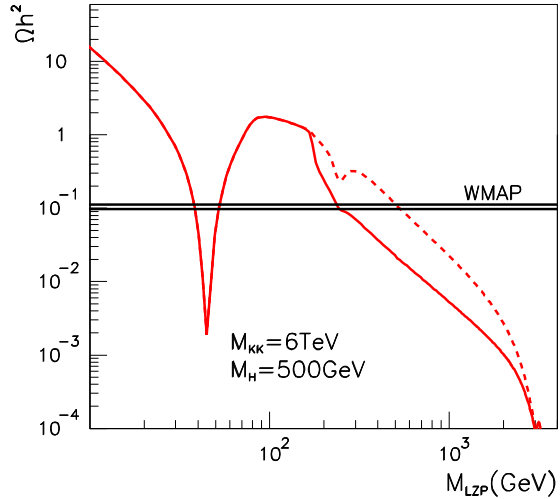, width=9.cm}}
  \caption{Relic density  of the  LZP for
  $M_H=500$~GeV, 
  $g_{10}=1.2,\cnul=0.1, r=0.1, \Lambda=2 M_{KK}$, corresponding to 
 $10^{-2}<g_Z<7 . 10^{-2}$ and   $\theta_X=3.2 \ 10^{-4}$,
 with (full)  and without (dash) the contribution of $X_S$. }
   \label{fig:noxs}
\end{figure}
%
%
We consider the range $-0.7<\cnur<0$ which leads to a
LZP  mass  ranging from
1 (2.4) GeV to 2 (4) TeV, for $M_{KK}$=3 (6)TeV
and $\Lambda=2M_{KK}$.
Figure~\ref{fig:omlzp} shows the behaviour of $\Omega
h^2$ as a function of the LZP mass for two  extreme values of the $SO(10)$
coupling, $\gten=0.3,1.2$ and $M_{KK}$= 3,6, 10 TeV. 
We see respectively the effects of the $Z$, the Higgs and $Z'$ resonances.
At large LZP mass, the
main annihilation channels are into $W$ pairs,  heavy fermions ($b,t$) 
 through $Z'$ exchange and  into top quarks through t-channel exchange of $X_S$.

\subsubsection {Annihilation into top}
Models with Pati-Salam $SU(4)\times
SU(2)\times SU(2)$ gauge group instead of $SO(10)$  share
most of the properties that we have discussed. As far as dark matter annihilation is concerned, the main difference comes from the absence of the  $X_S$
gauge boson. For this reason, we estimate separately the
contribution from the t-channel diagram with $X_S$ exchange. It is particularly important near the threshold where the s-channel
contribution of neutral gauge bosons is not so large. In
Fig.~\ref{fig:noxs}, we compare  $\Omega_{LZP} h^2$ with and
without the contribution from $X_S$.  The difference
can reach an order of magnitude. As one moves in the TeV region,
the shift is more modest as the
  annihilation into $W$ pairs becomes much more important. The net impact of ignoring
the contribution of $X_S$ would be to increase the lower bound  on
$\mlzp$ from 250 to 600 GeV for this choice of parameters.
\subsection{Coannihilation with KK fermions}

\begin{figure}[!htb]
  \centerline{\epsfig{file=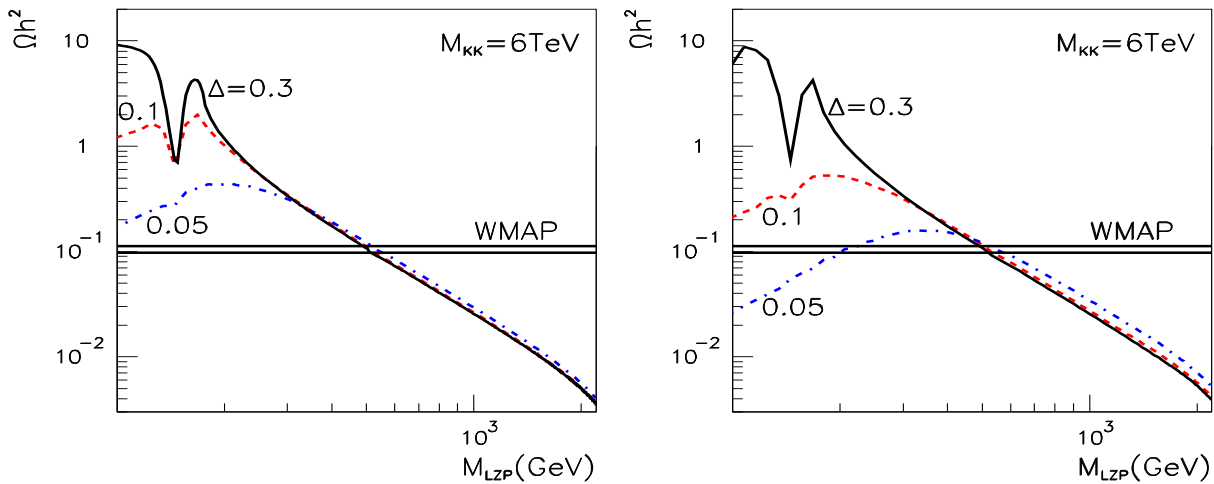, height=9.15cm,width=18cm}}
  \caption{LZP relic density  for
  $M_{KK}=6$~TeV,
  $g_{10}=0.78,\cnul=0.4, r=0.1,\Lambda=2M_{KK}$, corresponding to $5. 10^{-3}<g_Z<10^{-2}$ and $\theta_X=3.10^{-4}$, including the contribution of coannihilation
channels.  The NLZP is a) $\tau_R$  and  b) $(\tau_L, \nu_L)$.
The NLZP-LZP mass difference is fixed to $\Delta=0.3$ (black/full), $\Delta=0.1$ (red/dash) and 
$\Delta=0.05 \mbox{ (blue/dash-dot)}$. 
}
  \label{fig:coan}
\end{figure}
\begin{figure}[!htb]
  \centerline{\epsfig{file=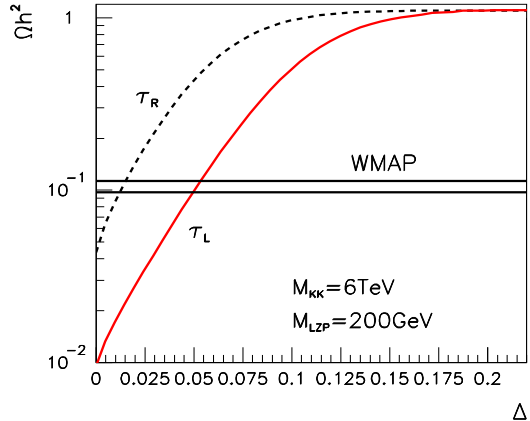, height=8.15cm,width=8.5cm}}
  \caption{LZP relic density versus the NLZP-LZP mass
  difference, $\Delta$, when  the NLZP is  $(\tau_L,\nu_L)$ (red),  $\tau_R$
(black).
  Here, $g_{10}=0.78$, $M_H=500$~GeV  $\cnur=-0.55$, $\cnul=0.4$ and
$c_f=0.1$ for all other KK fermions.}
\label{figure:delta}
\end{figure}
As we said, all KK fermions belonging to the multiplet containing $t_R$,  except the KK mode of $t_R$,  are expected to be
 light (compared to the KK gauge boson mass) and close to the LZP mass.
Among all possible coannihilation channels, we first examine those
with KK leptons. The couplings of KK leptons $\tau_L,\tau_R,\nu_L$
to the $Z$ are much enhanced compared to the LZP which is the only one to be suppressed by mixing angles. Therefore, we expect a large
impact of coannihilation channels for $\mlzp \approx 100$ GeV where
the $Z$ exchange dominates. On the other hand, the coupling of KK
leptons to the $Z'$ are only slightly suppressed  as compared to the
$Z'$-LZP coupling and
coannihilation effects are more modest when the $Z'$ exchange dominates  as illustrated in
Fig.~\ref{fig:coan}.
The coannihilation channels are dominated by the NLZP pair annihilation into fermion
pairs. Fig.~\ref{figure:delta}  shows the impact of coannihilation
for two sets of parameters
  where $\Omega_{LZP} h^2$ is much above WMAP
if one considers only annihilation processes. The mass of the LZP is fixed to 200 GeV. We vary
the mass of either the left-handed KK leptons or the right-handed lepton, $\tau_R$.
For left-handed leptons, the relic density drops within the WMAP range for $\Delta_L\approx 4\%$
while a much smaller mass splitting is needed for $\tau_R$ ($\Delta_R=1\%$).
\begin{figure}[!htb]
  \centerline{\epsfig{file=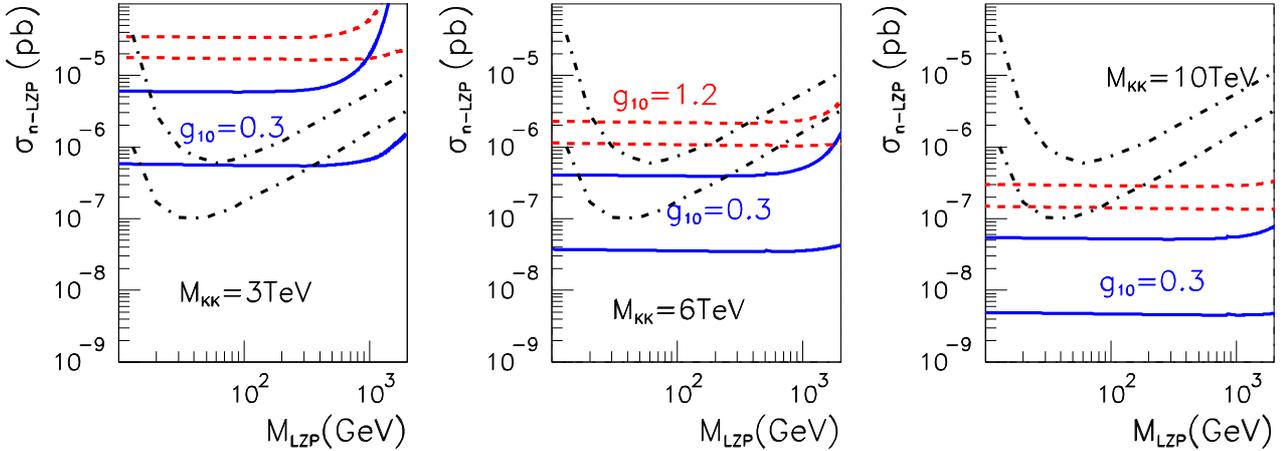, width=19cm}}
  \caption{LZP scattering on neutron for the same parameter values
as  Fig.~\ref{fig:omlzp}:
Solid blue and red dashed curves correspond respectively to $g_{10}=0.3$  and $g_{10}=1.2$. Lower (upper) curves are for $\cnul$=0.9 (-0.1).
The two dash-dot lines are the CDMS and Xenon constraints \cite{Akerib:2004fq,latest_xenon}.  
} \label{fig:ddlzp}
\end{figure}

We have also considered coannihilation channels with KK quarks, specifically $b_L,t_L$ since $b_R$
has no $Z_3$ charge. Coannihilation channels are dominated by the NLZP pair annihilation into fermions and gauge bosons, in
particular  channels involving quarks and gluons. Because of the strong coupling involved, one does not need such a small mass splitting
with the LZP as in the KK lepton case. For the parameters of Fig.~\ref{figure:delta}, a mass splitting $\Delta_Q=12\%$ is sufficient.
We ignore coannihilation with KK gauge bosons for several reasons. First, the motivation for studying the LZP model is that there is a strong rationale for the LZP to be significantly lighter than KK gauge bosons. Second, 
as we have seen above,  LZP's in the TeV range have already a very low  relic density, coannihilation would further reduce this relic density. Third, we made the symplifying assumption
that all KK gauge bosons are degenerate, this approximation is not appropriate in the case of coannihilation where
the mass difference with the LZP should be known precisely.
%
\begin{figure}[!htb]
  \centerline{\epsfig{file=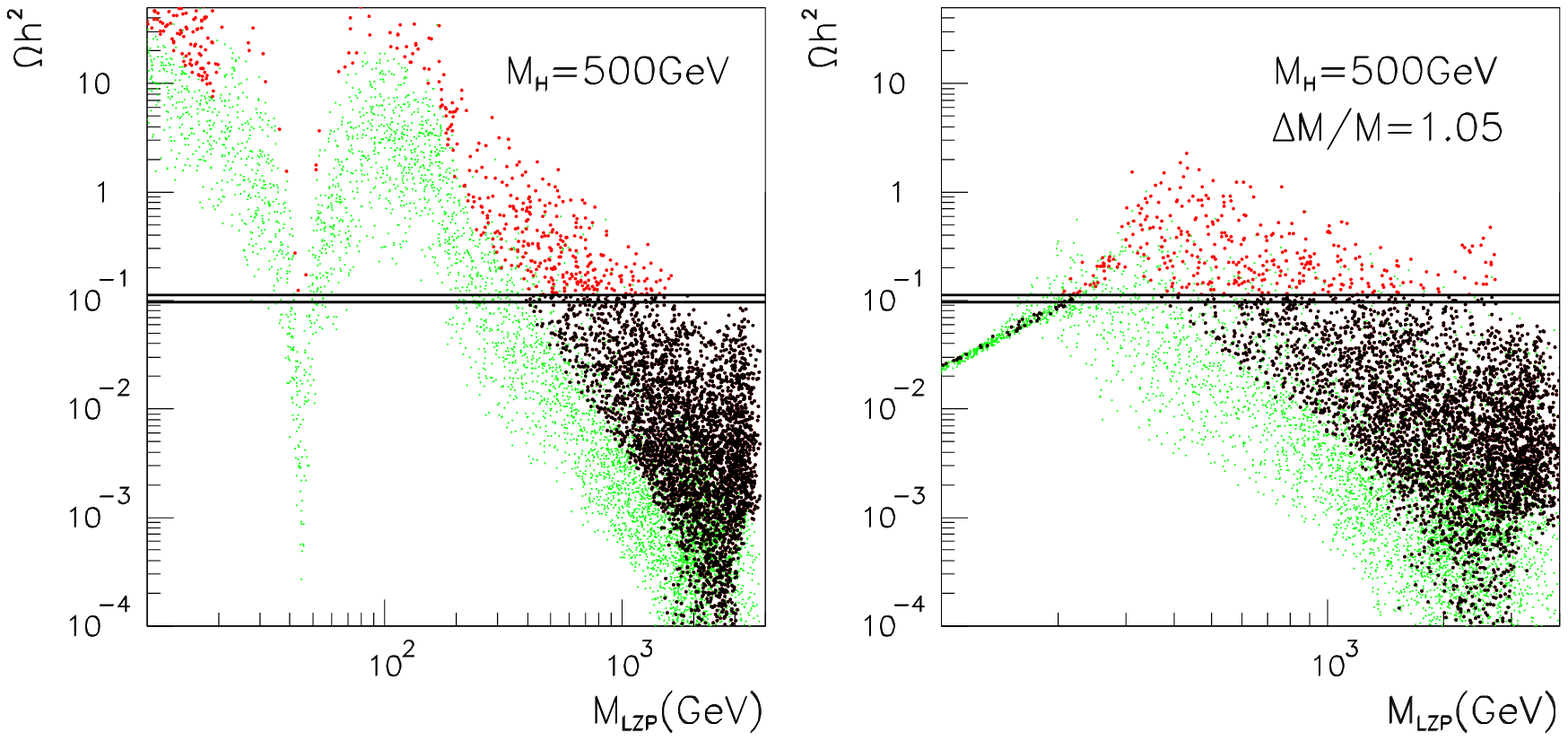, width=16cm}}
  \caption{a) Relic density  of LZP for a
  scan over input parameters, $3<M_{KK}<6$~TeV, $0.3<g_{10}<1.2,
  -0.1<\cnul<0.9$, $0<r<0.2,M_{KK}<\Lambda<2M_{KK}$. Here $c_i=0.1$ for all other KK
  fermions and $M_H=500$~GeV. Green/light grey dots show all models included in the scan, red/grey crosses
show only models that also satisfy the XENON bound while black crosses show models that satisfy
in addition the WMAP upper limit.  The WMAP preferred region corresponds to the
  horizontal lines.  
b) Same as above except that the masses of the KK leptons are fixed to $1.05\mlzp$. 
   } \label{fig:scan}
\end{figure}
\subsection{Direct detection and Summary}

As pointed out in ~\cite{Agashe:2004ci,Agashe:2004bm}, the LZP models have  large rates in direct detection.
In Figure~\ref{fig:ddlzp} we plot $\sigma_{n-LZP}$ as a function
of the LZP mass for the same range of parameters as in
Fig~\ref{fig:omlzp}, compared to the CDMS and XENON limits 
\cite{Akerib:2004fq,latest_xenon}. 
 For $M_{KK}=3$~TeV, the  $Z-Z'$ mixing is large and the LZP-nucleon cross-section is above
the XENON limit for most of the parameter space. 
From these figures, we conclude that it
might be difficult to satisfy both the WMAP upper bound on the
relic density and the limit from direct detection. 

To investigate
this more carefully, we  scan randomly,  generating  $10^5$
models within the parameter space $3\mbox{ TeV}<M_{KK}<6$~TeV, $.3<g_{10}<1.2$,
$-0.1<\cnul<0.9$ and  $-0.7<\cnur<0$. Here we fix $\Lambda=2M_{KK}$
and $r=0.1$. We keep  only models for which $\Omega_{LZP} h^2 <0.13$.
The predictions for $\sigma_{n-LZP}$ range from
$\sim 10^{-7} \ {\rm  to } \sim 10^{-4}$~pb. Most of these
values are already excluded by XENON, see Fig.~\ref{fig:scan},  in
particular for light LZP's where annihilation proceeds through $Z$
or Higgs exchange. Only models for which $\mlzp> 400$ GeV survive.
The full set of parameters that we have scanned all predict large
signals for direct detection and the whole parameter space, at
least for $\mlzp < 1$~TeV, will be probed within a few years as this
requires only a factor 6 improvement in the best limit.
If we had scanned up to KK mass of 10 TeV, the lower limit
on $\sigma_{n-LZP}$ would be roughly the same and we would reach
similar conclusions once we impose the WMAP constraint.

The presence of coannihilation channels changes the picture
since coannihilation helps reducing the relic density but does not
affect the elastic scattering cross section. Therefore, models
with weak couplings to the $Z$ that are still allowed by direct
detection experiments can also satisfy the upper limit on the
relic density. For example, $\mlzp=400$~GeV, $M_{Z'}=6$ TeV and $\cnul=0.4$ $(0.9)$ lead
respectively to $\sigma_N= 3.7$ $(2.6) 10^{-7}$~pb, below the
Xenon limit. 

\section{Acknowledgements}
This work was supported in part by GDRI-ACPP of CNRS. 
We are grateful to K. Agashe for  discussions.
We thank M. Cirelli and A. Boyarsky for useful comments.
A. Pukhov thanks the CERN Theory division for their hospitality. 

\section{Appendix}
Here we give the
details of the implementation of the LZP model \cite{Agashe:2004ci,Agashe:2004bm} within \calchep \ and \micro. 

\subsection{Particle Content}

We only consider the first level KK  fermions
that are in the same multiplet as the SM right-handed top quark, which has $B=-1/3$.
We ignore all other KK fermions of the third generation as well as those of the first and second generations.
Explicitly  we include
\begin{equation}
t_L,b_L,b_R, \tau_L, \nu_L, \tau_R, \nu_R=LZP
\end{equation}
where we use the shorthand notation $t_L=t_L^{(1)}$...  The
 first KK level of $t_R$ 
is not included. $t_R$ has 
different boundary conditions to provide the zero mode of the RH top quark and its first KK mode
is in the multi-TeV range. 
We include of course the zero mode of the right-handed
top, $t_R^{(0)}$. Although we do not include  $\tilde\nu'_L$ of the
$B=1/3$ multiplet, this particle plays
 a role because it mixes with the LZP via the top Yukawa coupling. This
 influences the coupling of the LZP
 to  the $Z$ and to the Higgs. We will use the $c$-parameter associated
 with it, $c_{\nu'_L}$. 
The 5D Yukawa coupling of the top  is related to the c-parameters of the zero
mode of the top quark,
\begin{equation}
k \lambda_5=\frac{1}{\sqrt{(1-2c_L)(1-2c_R)}}
\end{equation}
We choose $c_{L}=0.4$, $c_{R}=-0.5$ leading to $k \lambda_5=1.58$.  
\begin{table*}[htb]
\caption{\label{vertices} KK states included in the model}
\vspace{.5cm}
\begin{tabular}{|l|l|l||l|l|l||l|l|l|}
\hline
name&Charge&$Z_3$ & name&Charge& $Z_3$ &name&Charge&$Z_3$\\
\hline
$t_L$  & 2/3 & -2/3 & $\tau_L$  & -1 & -1/3 & $X_S$  & 2/3 & -1/3\\
$b_L$  & -1/3 & -2/3 & $\nu_L$  & 0 & -1/3 & $Z'$  & 0 & 0\\
$t_R \mbox{ (SM zero mode) }$  & 2/3 & 0 & $\tau_R$  & -1 & 1/3  & $W'$  & 1 & 0\\
$b_R$  & -1/3 & 0 & $\nu_R$  & 0 & 1/3 &  &  &  \\
\hline
\end{tabular}
\label{table:particles}
\end{table*}

The new gauge bosons of
$SU(2)_L\times SU(2)_R\times U(1)$ are 
$Z'$ and $W'$ (with dominantly
right-handed couplings to fermions).  
The two neutral gauge bosons mass eigenstates are a mixture of the
standard model $Z$ and the first KK mode of the $Z'$ ($\tilde
Z'$)
\begin{eqnarray}
Z&=&\cos\mixzzp Z_{SM}+\sin\mixzzp \tilde{Z'}\\
Z'&=&-\sin\mixzzp Z_{SM}+\cos\mixzzp \tilde{Z'}
\end{eqnarray}
Since the lighter state has to be dominantly the standard model
$Z$, $\sin\theta_{ZZ'}\ll1$,  and we will ignore the contribution of
terms that are suppressed by mixing except when the leading term
is absent. For charged gauge bosons, the lighter state is denoted
by $W$ and is dominantly the standard $W_L$ while the heavier
state $W'$ is dominantly $W_R$.
\begin{eqnarray}
W&=&\cos\theta_{WW'} W_L+\sin\theta_{WW'} W_R\\
W'&=&-\sin\theta_{WW'} W_L+\cos\theta_{WW'} W_R
\end{eqnarray}
 Again we ignore the contribution of terms suppressed by mixing,
 $\sin\theta_{WW'}<<1$.  The mixing angle depends on the overlap between the wave function of the Higgs and that of $Z'$ or $W'$, denoted ${\cal G}_{Z'}^H$ (see Eq.~\ref{eq:gzph})
 \begin{equation}
\theta_{WW'}=g_{10} \frac{s_W}{e} {\cal G}_{Z'}^H
\left(\frac{M_W}{M_{W'}}\right)^2
\end{equation}
\begin{equation}
\label{eq:thetaZZp}
\theta_{ZZ'}=\frac{g_{10}}{\sqrt{10}} \frac{2s_Wc_W}{e} {\cal G}_{Z'}^H
 \left(\frac{M_Z}{M_{Z'}}\right)^2
\end{equation}
In addition, we include the
leptoquark gauge boson of $SO(10)$, $X_s$, which is coloured and has electric charge
2/3. 
The mixing angle $\theta_X$  between the two $SO(10)$ 
gauge bosons $X'$ and $X_S$
 appears in the $\overline{t} \nu_L  {X}_s$ vertex and reads (see Ref.~\cite{Agashe:2004ci,Agashe:2004bm} for more details)
 \begin{equation}
 \theta_X= k \lambda_5(v/\Lambda)r^2
 \label{thetaXmixing}
 \end{equation}
Here $r$ parametrizes the amount of bulk GUT breaking and will be chosen $\sim 0.1$.
\subsection{Parameters}
We list the parameters of the model in Table~\ref{table:parameters}. Unless otherwise noted we take
$\mxs=M_{Z'}=M_{W'}=M_{KK}$.
\begin{table*}[!htb]
\caption{Parameters of the model} \vspace{.5cm}
\begin{tabular}{|l|l||l|l|}
\hline
Parameter&&Parameter& \\
\hline
$M_{KK}$ & Mass of KK gauge bosons  & $c_R$  & c-parameter of the RH top quark  \\
$r_c$  & $\log(M_{Pl}/M_{KK})\approx \log(10^{15})$ & $c_L$ & c-parameter of the LH top quark  \\
 $r$   & SO(10) bulk breaking & $c_{\nur}$  & c-parameter of the LZP  \\
$\Lambda$  & Cut-off of RS theory  & $c_{\tau_R}$ & c-parameter of  $\tau_R$ \\
$g_{10}$ & 4D SO(10) coupling  & $c_{\tau_L}$ & c-parameter of $\tau_L$ and $\nu_L$  \\
&     & $c_{t_L}$ &  c-parameter of  $t_L$ and $b_L$ \\
  $M_H$  & Higgs mass         & $c_{b_R}$ & c-parameter of $b_R$  \\
  &         & $c_{\nu'_L}$ & c-parameter of  $\nu'_L$  \\
\hline
\end{tabular}
\label{table:parameters}
\end{table*}
The masses $ M (c,M_{KK})$ in TeV of the KK fermions are approximated by (see Fig.~1 of Ref.~\cite{Agashe:2004bm}),
 \beqn M (c,M_{KK}) \times 2.405 &\approx  & {M_{KK}} \times 2
\sqrt{a(c)\left(a(c)+1\right)}
e^{- r_c a(c)}   \ \ \   \mbox{if } c < -\frac{1}{2}-\epsilon\nonumber\\
& & {M_{KK}}  \times 2\sqrt{c+\frac{1}{2}}    \ \ \    \mbox{if }
-\frac{1}{2}+\epsilon<c\leq-\frac{1}{4}\nonumber\\
 & &{M_{KK}}\times \frac{\pi}{2}({1+c})  \ \ \  \mbox{if }
c\gsim-\frac{1}{4}
\label{formula:mLZP}
\eeqn 
where $a(c)=|c+\frac{1}{2}|$.  The numerical value 2.405 arises from the solution to the eigen value problem in the Randall-Sundrum geometry where the wave functions are given by Bessel functions.

5D fermions lead to two chiral fermions in 4D. The two chiralities have different boundary conditions on the TeV and Planck branes. The LH helicity of the LZP turns out to be localized near the Planck brane while the RH helicity is near the TeV brane (see Ref.~\cite{Agashe:2004bm} for more details). Therefore, only the RH helicity couples significantly to the KK gauge bosons.
\subsection{Wave functions}
\subsubsection{Fermions: SM $t_R$ and KK $\nu_R$}
\beqn
 f_{t_R}(z,c)= (ze^{r_c} )^{2-c} \left[
\frac{e^{r_c}(1-2c)} {e^{r_c(1-2c)} -1} \right]^{1/2}
\eeqn 
\beqn 
f_{LZP}(z,c,m) =\left(ze^{r_c} \right)^{\frac{5}{2}}
\frac{e^{r_c/2}}{\sqrt{r_c} N(c,m)} \left[
J_{a(c)}(mz)+b(c,m) Y_{a(c)}(mz)\right] 
\eeqn
\beqn
\mbox{where } \ \ \ b(c,m) =-\frac{ J_{a(c)}(me^{-r_c})}{Y_{a(c)}(me^{-r_c})}
\eeqn
\beqn N^2(c,m)_{\stackrel{c<-1/2}{c\geq -1/2}} &=& \frac{e^{2r_c}}{2r_c}
\left[ \left(J_{a(c)}(m) + b(c,m) Y_{a(c)}(m)\right)^2 \right.\nonumber\\
 && \left. -e^{-2r_c} \left(J_{a(c)\pm 1}(me^{-r_c} ) + b(c,m)
Y_{a(c)\pm 1}(me^{-r_c} )\right)^2 \right] 
\eeqn 
\subsubsection{KK gauge bosons}
\beqn
f_{ga}(z,x)&=& e^{r_c/2} \frac{z}{N_{ga}(x)} \left[J_1(zx)+b_{ga}(x) Y_1(zx)
\right]\\
b_{ga}(x)&=&-\frac{J_1(xe^{-r_c})}{Y_1(xe^{-r_c})}\\
N^2_{ga}(x)&=&\frac{1}{2} \left[ \left(J_1(x)+b_{ga}(x)Y_1(x)\right)^2- e^{-2r_c}
\left( J_0(x e^{-r_c})+b_{ga}(x)Y_0(x e^{-r_c})\right)^2\right]\nonumber
\eeqn
\subsubsection{Higgs boson} \beqn f_h(z)=\sqrt{\frac{2z^2
e^{r_c}}{1-e^{-2r_c}}} \eeqn
\subsubsection{Wave function overlaps}
\begin{eqnarray}
g_1^{f}(c_1,c_2)& =&  e^{-r_c/2} \int_{e^{-r_c}}^1 dz  \left(z
e^{r_c}\right)^{-4}
f_{ga}(z,2.405) f_{t_R}(z,c_2) f_{LZP}(z,c_1,2.405)\;\;\;\; \;\;\;\; \;\;\;\;\; \;\;\;\;\;\;\;\;\;\;\\
 g_{2}^{f}(c_1,c_2)&=&e^{-r_c/2} \int_{e^{-r_c}}^1 dz
\left(ze^{r_c}\right)^{-4}
 f_{ga}(z,2.405)
f_{LZP}(z,c_1,2.405) f_{LZP}(z,c_2,2.405) \\
 g_{Z'}^{l}(c)&=&e^{-r_c/2} \int_{e^{-r_c}}^1 dz
\left(ze^{r_c}\right)^{-4}
 f_{ga}(z,2.405)
f^2_{LZP}(z,c,2.405)\\
g_{Z'}^{t}(c)&=&e^{-r_c/2} \int_{e^{-r_c}}^1 dz
\left(ze^{r_c}\right)^{-4} f_{ga}(z,2.405) f_{t_R}^2(z,c) \\
{\cal G}_{Z'}^H &=&\sqrt{r_c e^{-r_c}}\int_{e^{-r_c}}^1 dz   (z e^{r_c})^{-1}
f_{ga}(z,2.405) f_h^2(z) 
\label{eq:gzph}
\end{eqnarray}
%
%
\begin{table*}[!htb]
\caption{\label{vertices} Couplings of KK fermions  to neutral SM
gauge bosons} \vspace{.5cm}
\begin{tabular}{|l|l|l|l|}
\hline
Vertex&Value&Vertex&Value\\
\hline
$\overline{t_L} t_L G_\mu$  &$ g_s\gamma_\mu$  &$\overline{\tau_R} \tau_R A_\mu$   &$ -e\gamma_\mu$ \\
$\overline{t_L} t_L A_\mu$   &$ \frac{2}{3}e\gamma_\mu$  & $\overline{\tau_R} \tau_R Z_\mu$   &$ \frac{e\sw}{\cw}\gamma_\mu$ \\
$\overline{t_L} t_L Z_\mu$   &$ \frac{e}{\cw\sw}\left(\frac{1}{2}-\frac{2}{3}\sw^2\right)\gamma_\mu$ &  $\overline{\tau_L} \tau_L A_\mu$   &$ -e\gamma_\mu$ \\
$\overline{b_L} b_L G_\mu$   &$ g_s\gamma_\mu$ & $\overline{\tau_L} \tau_L Z_\mu$   &$ -\frac{e}{2\cw\sw}(1-2\sw^2)\gamma_\mu$ \\
$\overline{b_L} b_L A_\mu$   &$ -\frac{1}{3}e\gamma_\mu$ & $\overline{\nu_L} \nu_L Z_\mu$   &$ \frac{e}{2\cw\sw}\gamma_\mu$ \\
$\overline{b_L} b_L Z_\mu$   &$
\frac{e}{\cw\sw}\left(-\frac{1}{2}+\frac{1}{3}\sw^2\right)\gamma_\mu$
& $\overline{\nu}_R \nu_R Z_\mu$
 &$ \frac{1}{2}g_Z^{\nu_R}\gamma_\mu(1+\gamma_5)$ \\
$\overline{b_R} b_R G_\mu$   &$ g_s\gamma_\mu$ &&\\
$\overline{b_R} b_R A_\mu$   &$ -\frac{e}{3}\gamma_\mu$ &&\\
$\overline{b_R} b_R Z_\mu$   &$ \frac{e\sw}{3\cw}\gamma_\mu$
&&\\
$\overline{b_R} b Z_\mu$   &$ -\frac{e}{2\sw\cw}
f_{c_R}\frac{m_t}{M_{b_R}}\gamma_\mu(1-\gamma_5)$ &&  \\ 
\hline
\end{tabular}
\end{table*}
\begin{table*}[!htb]
\caption{\label{verticesW} Couplings of KK fermions to $W^\pm$}
\vspace{.5cm}
\begin{tabular}{|l|l|l|}
\hline
Vertex&Value& \\
\hline
$\overline{b_R} t W^-_\mu$   &$ \frac{1}{2} (g_{Rq}\theta_{WW'} \gamma_\mu(1+\gamma_5)+\frac{e f_{c_R} m_t}{\sqrt{2}M_{b_R}\sw}\gamma_\mu(1-\gamma_5))$& $g_{R_q} =g_{10} \sqrt{\frac{r_c}{2}}g_{1f}(c_{b_R},c_R)$\\
$\overline{b_L} t_L W^-_\mu$  &$ \frac{e}{\sqrt{2}\sw} V_{tb}\gamma_\mu$ & \\
$\overline{\tau_L} \nu_L W^-_\mu$   &$ \frac{e}{\sqrt{2}\sw}\gamma_\mu$ &\\
$\overline{\tau_R} \nu_R W^-_\mu$   &$ \frac{1}{2}g_{R_l}\theta_{WW'}\gamma_\mu(1+\gamma_5)$ &$g_{R_l} = g_{10} \sqrt{\frac{r_c}{2}} g_{2f}(\cnur,c_{\tau_R})$\\
\hline
\end{tabular}
\end{table*}

\begin{table*}[!htb]
\caption{\label{verticesZ'} Couplings of SM particles and KK fermions to new gauge
bosons} \vspace{.5cm}
\begin{tabular}{|l|l|l|}
\hline
Vertex&Value&\\
\hline
$\overline{t} \nu_R {{X}_s}$   &$ \frac{1}{2}g_1\gamma_\mu(1+\gamma_5)$ &$g_{1} = g_{10} \sqrt{\frac{r_c}{2}} g_{1f}(\cnur,c_R)$\\
$\overline{t} \nu_L  {X}_s$   &$ \frac{g_4}{2} \theta_X\gamma_\mu(1-\gamma_5)$ & $g_{4} = g_{10} \sqrt{\frac{r_c}{2}} g_{1f}(\cnul,\cnur)$\\
$\overline{\nu}_R \nu_R Z'_\mu$   &$ \frac{g_{Z'}}{2}\gamma_\mu(1+\gamma_5)$ & $g_{Z'} = g_{10} \sqrt{\frac{5 r_c}{8}} g_{Z'}^{l}(\cnur)$\\
$\overline{\tau}_R \tau_R Z'_\mu$   &$ \frac{g_{\tau}}{2}\gamma_\mu(1+\gamma_5)$ &
$g_{\tau} = g_{10} \sqrt{\frac{r_c}{40}} g_{Z'}^{l}(\ctau)$\\
$\overline{t} t Z'_\mu$   &$ \frac{g_{tR}}{2}
\gamma_\mu(1+\gamma_5)+ \frac{g_{tL}}{2}\gamma_\mu(1-\gamma_5)$ &
$g_{tR} = g_{10} \sqrt{\frac{r_c}{40}} g_{Z'}^{t}(c_R)$\\
$\overline{b} b Z'_\mu$   &$ \frac{g_{tL}}{2}\gamma_\mu(1-\gamma_5)$ & $g_{tL} = -g_{10} \sqrt{\frac{r_c}{40}} g_{Z'}^{t}(c_L)$\\
$H Z_\mu Z'_\nu$ &$\frac{e v g_{HZ}}{\sw\cw} g_{\mu\nu}$& $g_{HZ}=g_{10}\sqrt{\frac{r_c}{40}}$\\
$\overline{\nu}_L \nu_L
Z'_\mu$&$\frac{g_{Z'}^L}{2}\gamma_\mu(1-\gamma_5)$ &
{${g_{Z'}^L}=3g_{10}\sqrt{\frac{r_c}{40}}g_{Z'}^{l}(\ctaul)$ } \\
$\overline{\tau}_L \tau_L Z'_\mu$   &$ \frac{g_{Z'}^L}{2}\gamma_\mu(1-\gamma_5)$ &\\
$\overline{t}_L t_L Z'_\mu$   &$
\frac{g_{Z'}^t}{2}\gamma_\mu(1-\gamma_5)$
&${g_{Z'}^t}=-g_{10}\sqrt{\frac{r_c}{40}} g_{Z'}^{l}(\ctl)$\\
$\overline{b}_L b_L Z'_\mu$ &
$\frac{g_{Z'}^t}{2}\gamma_\mu(1-\gamma_5)$
&\\
$\overline{b}_R b_R Z'_\mu$   &$
\frac{g_{Z'}^b}{2}\gamma_\mu(1+\gamma_5)$
&${g_{Z'}^b}=-3 g_{10}\sqrt{\frac{r_c}{40}} g_{Z'}^{l}(\cbr)$\\
$W^+_\alpha W^-_\beta Z'_\mu$   & $-g c_W\mixzzp
[(p_2^\alpha-p_3^\alpha) g^{\beta\mu}+
(p_3^\beta-p_1^\beta) g^{\alpha\mu}+(p_1^\mu-p_2^\mu) g^{\alpha\beta}]$  & \\
$\overline{b_R} t W'^-$ & $\frac{g_{Rq}}{2}\gamma_\mu(1+\gamma_5)$&  \\
$\overline{b_L} t_L W'^-$ & $\frac{e}{\sqrt{2}s_W}V_{tb} (-\theta_{WW'})\gamma_\mu$&  \\
$\overline{b} t W'^-$ & $\frac{e}{2\sqrt{2}s_W} (-\theta_{WW'})\gamma_\mu(1-\gamma_5)$& \\
$\overline{\tau_R} \nu_R W'^-$ & $\frac{g_{Rl}}{2}\gamma_\mu(1+\gamma_5)$&   \\
$\overline{\tau_L} \nu_L W'^-$ & $\frac{e}{\sqrt{2}s_W}
(-\theta_{WW'})\gamma_\mu$&
 from $W_L-W_R$ mixing \\
$\overline{\tau} \nu W'^-$ & $\frac{e}{\sqrt{2}s_W}
(-\theta_{WW'})\gamma_\mu(1-\gamma_5)$&
 from $W_L-W_R$ mixing \\
$W'^+_\alpha W^-_\beta Z$   &   $-g c_W\theta_{WW'}
[(p_2^\alpha-p_3^\alpha) g^{\beta\mu}+
(p_3^\beta-p_1^\beta) g^{\alpha\mu}+(p_1^\mu-p_2^\mu) g^{\alpha\beta}]$&  \\
$W'^+_\alpha W^-_\beta H$   & $g M_W \theta_{WW'} g^{\alpha\beta}$&  \\
\hline
\end{tabular}
\label{tab:couplingnewgaugebosons}
\end{table*}
\subsection{Neutrino mixing and coupling to the Higgs}

The large Yukawa coupling between the {\bf 16}  with $B=1/3$ containing  $t^0_L$ and the {\bf 16} with $B=-1/3$ containing $t^0_R$  generates a mixing mass term between ${\nu}'_L$ and the LZP. This induces LZP couplings to the Higgs and the $Z$ after diagonalization of the
 mass matrix
\begin{equation}
M_\nu=\left(
\begin{array}{cc}
\mlzp&m_{\nu_R\nu'_{L}}\\
0&m_{\nu'_{L}}
\end{array}
\right)
\end{equation}
\begin{equation}
\mbox{where }
\mlzp=M(\cnur,M_{KK}) , \;\;\;\;
m_{\nu'_{L}}=M(\cnul,M_{KK}) , \;\;\;\; 
m_{\nu_R\nu'_{L}}= m_t f_{c_L} f_{c_R} \xi_{\nu_R}.
\nonumber
\end{equation}
{\begin{equation}
\xi_f=1 \;\;\; \mbox{if}\;\;\; c_f>-1/2\;\;\; \mbox{and}\;\;\; \xi_f=1/f_{c_f}\;\;\; \mbox{if}\;\;\; c_f<-1/2  \ \ \mbox{and } f_{c}=\sqrt{\frac{2}{1-2c}}
\end{equation} 
The LH and RH helicities of the LZP (respectively 
the mass eigenstates of  $M^\dagger_\nu M_\nu$ and  $M_\nu M^\dagger_\nu $) are
\begin{eqnarray}
(\nu_1)_L&=&\cos\theta_L \hat{\nu}_R +\sin\theta_L \nu'_{L}\\
(\nu_1)_R&=&\cos\theta_R  \nu_R +\sin\theta_R \hat{\nu}'_{L}
\end{eqnarray}
where  $\hat{\nu}_R$  and $\hat{\nu}'_{L}$ are the 5D KK partners of  ${\nu}_R$  and ${\nu}'_{L}$, which have respectively LH and RH chiralities (see Ref.~\cite{Agashe:2004bm}  for more details).
Therefore,  the LZP coupling to the Higgs is given by:
\begin{eqnarray}
g_{H}= f_{c_L} f_{c_R} \sin\theta_L \cos\theta_R
\xi_{\nu_R} \label{hnunu}
\end{eqnarray}
 where we used
$2 k \lambda_5=f_{c_L}f_{c_R}$.
All fermions whose coupling to the Higgs are also induced via mixing
with a heavier KK state ($\tau_R,\tau_L,\nu_L,t_L,b_L$) have a coupling $g_{Hff}$  
 given by  Eq.~\ref{hnunu}. However,
in the mixing angle one should substitute the following masses for
the $\mbox{fermion } f$
\begin{equation}
m_1=M_{KK}(c_{f},\mxs)\;\;\; m_2=M_{KK}(\cnul,\mxs) \;\;\; m_3=
m_t f_{c_L} f_{c_R} \xi_{f}
\end{equation}
Here we assume that all the heavy states which mix
with level one KK fermion have a common mass equal to
$M_{KK}(\cnul,\mxs)$. 
The mixing angles are
\begin{equation}
\sin\theta_{L/R}=\left(1+\frac{(-m_1^2+m_2^2-m_3^2 + \sqrt{\Delta})^2}{4
m_{1/2}^2 m_3^2} \right)^{-1/2} \label{eq:neutrinomix}
\end{equation}
\begin{equation}
\cos\theta_R=- \frac{(-m_1^2+m_2^2-m_3^2+\sqrt{\Delta})}{2 m_2
m_3} \left(1+\frac{(-m_1^2+m_2^2-m_3^2+\sqrt{\Delta})^2}{4 m_2^2
m_3^2} \right)^{-1/2}
\end{equation}
 where $m_1=M_{LZP}$, $m_2=m_{\nu'_{L}}$, $m_3=m_{\nu_R\nu'_{L}}$ and $\Delta=-4 m_1^2 m_2^2 +(m_1^2+m_2^2+m_3^2)^2$.
\subsection{The $Z$-LZP   coupling}
The $Z$-LZP coupling ($g_Z$) has one component due to $Z-Z'$
mixing, $g_{ZZ'}^{\nu_R}$, as well as another component from
$\nu'_L\nu_R$ mixing, $g_{\nu_L\nu_R}^{\nu_R}$. Both components are
comparable although small. Explicitly,
 \begin{eqnarray}
  g_Z^{\nu_R}&= &g_{ZZ'}^{\nu_R}+  g_{\nu_L\nu_R}^{\nu_R} \nonumber\\
  g_{ZZ'}^{\nu_R}&=&g_{Z'}\ \theta_{ZZ'}\nonumber\\
 g_{\nu_L\nu_R}^{\nu_R}&=& g_Z^\nu \;\;\sin^2 \theta_{R}
 \label{Zn4n4} \end{eqnarray}
  where
$g_Z^\nu=e/2s_W c_W$ is the $Z$ coupling to $\nu'_L$ and  $\sin \theta_{L/R}$
is defined in Eq.\ref{eq:neutrinomix}. For $m_{\nu_R}\ll m_{\nu'_L}$, $\sin \theta_{L}\ll \sin \theta_{R}$ and we can neglect the coupling of the LH helicity of the LZP.
$\theta_{ZZ'}$ is given by Eq.~\ref{eq:thetaZZp} and $g_{Z'}$
is the $Z'$ coupling to the LZP, (see Table~\ref{tab:couplingnewgaugebosons}), 
\begin{equation}
g_{Z'}=g_{10} \sqrt{\frac{5r_c}{8}} g_{Z'}^{l}(c_{\nu_R})
\end{equation}


\end{document}